\documentclass[twocolumn,aps,prx,longbibliography]{revtex4-1}
\usepackage[utf8]{inputenc}
\usepackage[english]{babel}
\usepackage{amsmath,amssymb}
\usepackage{pifont}
\newcommand{\plaind}{\mathrm{d}}
\newcommand{\dbar}{\plaind\mkern-7mu\mathchar'26}

\newcommand{\deltabar}{\delta\mkern-8mu\mathchar'26}

\usepackage{tikz}
\usetikzlibrary{arrows,decorations.pathmorphing,backgrounds,positioning,fit,petri,arrows.meta}
\usepackage{pgfplots}
\usepackage{wasysym}

\begin{document}
\title{From Neuronal Spikes to Avalanches --  Effects and Circumvention of Time Binning.}
\author{Johannes Pausch}
\affiliation{Department of Applied Mathematics and Theoretical Physics, University of Cambridge, and St. Catharine's College, Cambridge, UK}
\date{\today}
\begin{abstract}
Branching with immigration is one of the most common models for the stochastic processes observed in neuronal circuits. However, it is not observed directly and,  in order to create branching-like processes, the observed spike time series is processed by attaching time bins to spikes. It has been shown that results such as criticality and size distributions depend on the chosen time bin. A different methodology whose results do not depend on the choice of time bin might therefore be useful and is proposed in this article.

The new methodology circumvents using time bins altogether by replacing the previously used discrete-time models by continuous-time models. First, the article introduces and characterises a continuous-time version of the branching process with immigration, which will be called pumped branching process, and second, it presents an \textit{analytical derivation of the corresponding spike statistics}, which can be directly compared to observed spike time series. The presented approach allows determining the degree of criticality, the average number of overlapping avalanches, and other observables without using a time bin. Furthermore, the effects caused by using time bins are analyzed and the influence of temporal and spatial subsampling discussed, all of which is compared to experimental data and supported by Monte Carlo simulations.
\end{abstract}

\maketitle 

\section{Introduction}
Spike trains recorded with electrode arrays are commonly used to study the functioning of the brain on a mesoscopic level.\cite{Hodgkin1946,Stein1965,Pfeiffer1965,Lamarre1971,Rinzel1983,Softky1993,Dayan2001,Beggs2003,Beggs2004,Beggs2008,Priesemann2009,Priesemann2013,Wilting2018a,Wilting2018b,Wilting2019,Wilting2019b,Miller2019} They offer insight into its activity while subjects are performing tasks or are in different stages of sleep.\cite{Lamarre1971,Priesemann2013} Whether the spiking is relatively regular or random, i.e. \textit{irregular}, continues to be a topic of scientific discussion.\cite{Lamarre1971,Tuckwell1979,Rinzel1983,Chrisodoulou2001,Moreno-Bote2014,Lengler2017,Mijatovic2018} A standard measure of irregularity is the coefficient of variation $c_V$ of the inter-spike interval (ISI). With $c_V$ equal to 1, the Poisson process divides the landscape of stochastic time series into more regular processes with small fluctuations and $c_V<1$ on the one hand, and on the other hand, highly irregular ones with large fluctuations and $c_V>1$. More importantly given data, the coefficient of variation can be used to identify or rule out candidates for the underlying stochastic process that governs neuronal spiking.

\begin{figure}[h!]
\begin{center}
\includegraphics[width=\columnwidth]{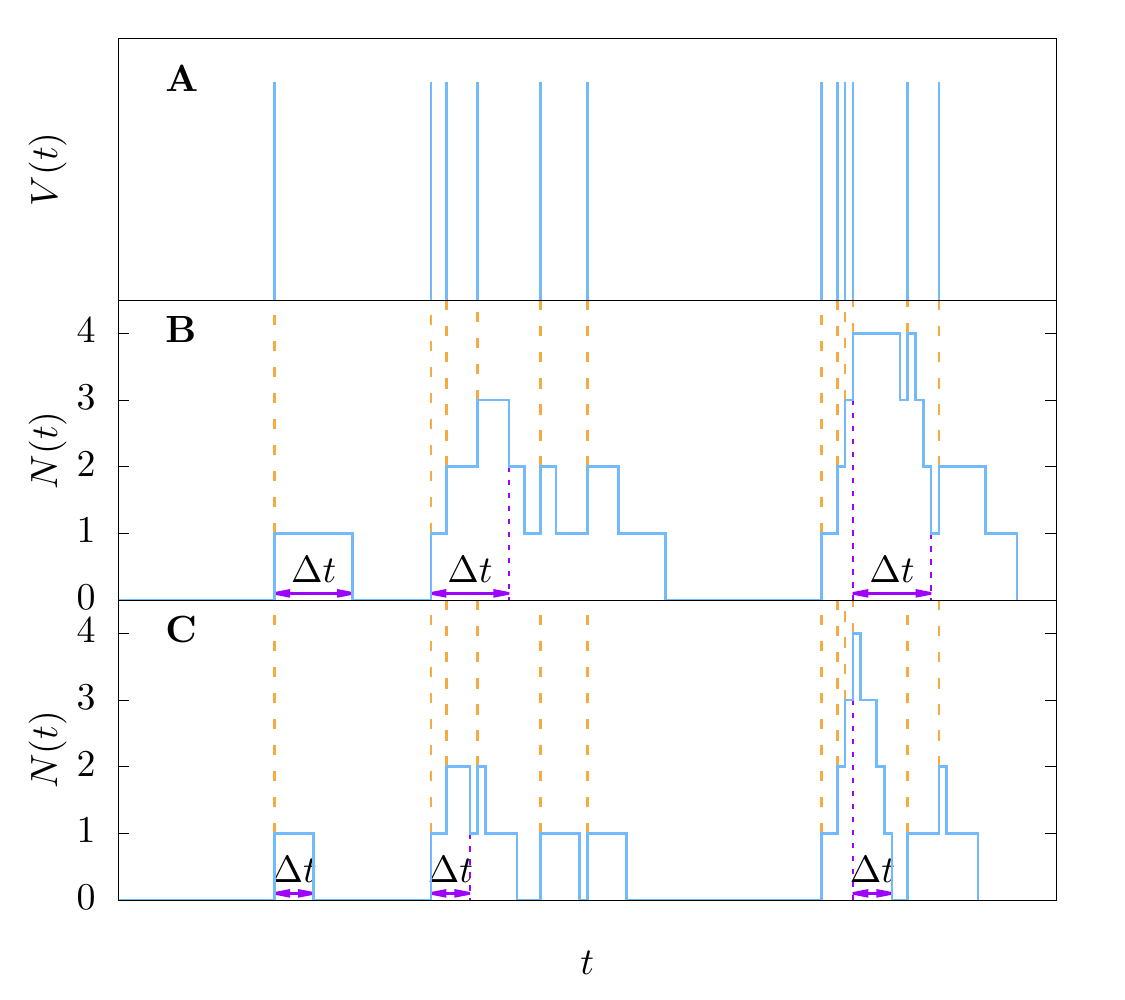}
\caption{Schematic transformation of a spike time series into a branching-like process. \textbf{A}: Spike time series. Spikes are typically identified as peaks in voltage $V(t)$ at implanted electrode arrays.\cite{Beggs2003} \textbf{B} and \textbf{C}: Resulting branching-like process with particle number $N(t)$.  A time bin $\Delta t$ is attached to each spike, representing a particle. Overlapping time intervals implicate that there are several particles present. A larger $\Delta t$ was used in panel B compared to panel C. Orange dashed lines are visual guides to see that spikes induce an increase in particle number. Purple arrows indicate three examples of bin sizes for panel B and C. Avalanches are defined as spells of activity, i.e. $N(t)>0$, between periods of inactivity where $N(t)=0$. Choosing different $\Delta t$ can lead to different activity, different avalanches and also changes their statistical properties.\cite{Priesemann2009}}
\label{fig-schematic}
\end{center}
\end{figure}

While the spiking of a single neuron might appear like a Poisson process, collectively, neurons do not spike independently.\cite{Beggs2004} The occurrence of spikes is expected to follow a branching structure just as the physical network of neuronal circuits on which they propagate.\cite{Beggs2003,Beggs2004,Zierenberg2020} The common way to generate a branching-like process from a spike time series is to attach a time bin of length $\Delta t$ to each spike.\cite{Beggs2003,Beggs2004,Priesemann2009,Priesemann2013} The number of overlapping time bins at a time $t$ is then interpreted as number of particles $N(t)$ in a branching process present at that time, Fig.~\ref{fig-schematic}. Time series generated in this way will be called `branching-like' because, a priori it is not clear whether such a time-series can mathematically be a branching process. A change in the chosen time bin $\Delta t$ generates a different branching-like process with different statistical properties.\cite{Priesemann2013}  The statistics of interest are those which show whether the system is close to \textit{criticality}, which is an important feature associated with divergence of correlation lengths \cite{Wilting2019,LeBellac1991,Garcia-Millan2018} and optimization of information processing.\cite{Beggs2008} However, it has been unknown so far if and how critical behavior of an underlying branching process could be identified directly from the observed spike time series. 

Criticality of branching processes can be identified by studying avalanches which are defined as spells of uninterrupted activity, i.e. spells where $N(t)>0$ interrupted by periods when $N(t)=0$. The statistics of avalanche sizes has been the most widely used observable for identifying critical behavior because its distribution follows a \textit{power law} if the system is at criticality.\cite{Paczuski1996,Garcia-Millan2018,Munoz2017} In practice, this feature is extremely difficult to test because tails of power laws or exponential distributions represent rare events. Sampling these rare events in simulations or experiments is challenging and comes with considerable uncertainties.\cite{Goldstein2004,Priesemann2018} Furthermore, subcritical branching processes can also show a power-law distribution of avalanche sizes with the same exponent over several magnitudes before being cutoff by an exponential tail. \cite{Garcia-Millan2018}

Recently, a new method has been established that allows estimating the closeness of the system to the critical point using auto-correlation functions.\cite{Wilting2018a,Wilting2018b,Wilting2019} This method is stable under subsampling of a mean-field neuronal network and it allowed identifying a regime close to criticality, called \textit{reverberating regime}, in which neuronal activity takes place. 

However, many of  these analyses \cite{Beggs2003,Beggs2004,Beggs2008,Priesemann2009,Lombardi2012,Lombardi2014,Priesemann2013,Wilting2018a,Wilting2018b,Wilting2019} build on several (but not necessarily all) underlying assumptions addressed here, which include: \begin{enumerate}\item The time bins $\Delta t$ are assumed to be fixed and equal for every spike despite experimental evidence and detailed modelling suggesting a spread of bin times \cite{Hodgkin1946,Stein1965,Dayan2001,Moreno-Bote2014}, see Fig.~\ref{fig-EXP-inter-spike-dsitribution}. In this article, bin times are distributed according to an exponential distribution, which allows implementing a continuous-time branching process, Sec.~\ref{sec-model}. \item While branching-like processes are created readily out of spike time series, it is unclear whether the resulting process would occur `naturally', e.g. in a Monte Carlo simulation of branching. This article presents phase space boundaries for the spike time series, outside which a `naturally occurring' branching process cannot correspond to the observed spiking. This means that the probability is zero for a spike time series to lie outside these boundaries if it is derived from a branching process, Sec.~\ref{sec-moment-ratio-map}. \item Furthermore, it has been unclear so far whether time series with equal fluctuations, as characterized by the coefficient of variation $c_V$, can result in branching processes with significantly different criticality measure. This article provides a map showing how two  differently critical branching processes can correspond to time series with the same coefficient of variation $c_V$, Sec.~\ref{sec-CV}. \item Typically $\Delta t$ is taken to be the average inter-spike interval $\langle\text{ISI}\rangle$, although some uncertainty about this choice has been published.\cite{Beggs2003,Beggs2004,Plenz2009,Priesemann2013} In this article the analytical relation between $\langle\text{ISI}\rangle$ and the average time for a particle to go extinct is presented. The avalanche size distribution is determined based on moments of the ISI rather than an unknown $\Delta t$, Sec.~\ref{sec-ISI-extinction-time}. \item This article points out a problem with the use of auto-correlations as tool to determine how close the system is to criticality \cite{Wilting2018b}: in continuous-time systems, their exponential decay depends not only on  the criticality parameter but its  product with a time scale, which cannot be estimated separately from auto-correlations alone, Sec.~\ref{sec-twotimecorrelation}.
\item Many analyses (but not all \cite{Munoz-Plenz2017,Wilting2018a,Wilting2018b,Wilting2019}) assume a separation of time scales, i.e.~that separately initiated neuronal avalanches are well separated in time within the local measurement. This allows comparing data to well-established results from self-organized criticality \cite{Beggs2003,Priesemann2009,Pruessner2012} and pure branching processes.\cite{Harris1964} However, based on experimental \textit{in vivo} data \cite{Buzsaki2009}, this article argues that overlaps can be very common, Sec.~\ref{sec-moment-ratio-map}.
\end{enumerate}


The aim of the article is to investigate the common approach of creating time series that look like branching processes from spike time series. The investigation is done by reversing the procedure: starting from continuous-time branching processes with immigration, spike time series are derived analytically. Thus, shortfalls of the approach are identified and an alternative methodology suggested. 

Data analysis is always an interpretation of data in the space of chosen models. This article suggests to change the space of models which consists of discrete-time branching processes with immigration by the model space of continuous-time pumped branching processes. There are two main conceptual arguments for this change:\begin{enumerate}\item the former relied on a choice of a fixed time bin which is difficult to justify from first principles, while the latter replaces the fixed time bins by a distribution for the ISI. The chosen ISI distribution might not be the best one, but given that continuous ISI distributions are observed and that many of them appear to have an exponential character, the analysis and use of the latter model space is hopefully regarded as useful by the research community. The continuous nature of the ISI distribution is shown for two example data sets in Fig.~\ref{fig-EXP-inter-spike-dsitribution};
\item the former model space does not have an intrinsic way of determining the best/correct time bin, while the latter intrinsically allows determining the full mapping from spikes to the model space, including its time scale.\end{enumerate} 
The resulting consequences will require careful neuroscientific interpretation which is beyond the scope of this article.

The structure of this article is as follows: First, the pumped branching process is introduced and characterized as a continuous-time version of the branching process with immigration. This part lays the groundwork for the second part and main focus of this article: the derivation of the \textit{induced spike statistics} and their application and comparison to experimental data. The pumped branching process is set up in Sec.~\ref{sec-model}, its steady state properties, including moments, probability distribution and correlation functions are derived in Sec.~\ref{sec-steady-state}, followed by the system's relaxation properties in Sec.~\ref{sec-relaxation}, which are important in connection with Monte Carlo simulations. In Sec.~\ref{sec-experiment}, the connection to experimental data is made by deriving the relation of the pumped branching process to spike time series in Sec.~\ref{sec-spike-time-distribution}. It is used to derive coefficient of variation, Sec.~\ref{sec-CV}, as well as a more useful moment-ratio map, Sec.~\ref{sec-moment-ratio-map}, which is the central object for identifying how close neuronal activity is to critical branching. The connection of the ISI to the expected particle extinction time and avalanche duration and size statistics are described in Sec.~\ref{sec-ISI-extinction-time} and subsampling effects are discussed in Sec.~\ref{sec-subsampling}. Finally, the article is concluded in Sec.~\ref{sec-conclusion}. Throughout, analytical results are verified with Monte Carlo simulations.

\section{Pumped branching}
\subsection{Model}\label{sec-model}
The following model is a continuous-time version of the branching process with immigration. It intends to imitate signal creation and propagation in the brain by representing the number of signals which are being propagated in a neuronal circuit as a particle number $N$, which is also called the state of the system. The input of a signal into a circuit is modelled by a spontaneous particle creation process with rate $\gamma$ (\textit{'pumping'}). When a signal spreads from one neuron to another, particles branch and create $K$ particles according to an offspring distribution $p_k$. At a branching event, the probability that the particle becomes $K$ particles equals $p_k$. However, neurons do not always spread signals they receive. They also inhibit the signal propagation, which is modelled implicitely as particle extinction and is incorporated in the offspring distribution by $p_0$. The reason why extinction is effectively inhibiting the future activity is that when a particle goes extinct, it is removed from the system. This removal implies a decreased branching activity because branching is proportional to the number of particles present in the system. How close the branching process is to criticality is determined by $\mathbb{E}[K]$, the expected number of offspring. The process is subcritical if $\mathbb{E}[K]<1$, it is supercritical if $\mathbb{E}[K]>1$ and it is critical for $\mathbb{E}[K]=1$. Many authors name $\mathbb{E}[K]=m$ or $\mathbb{E}[K]=\sigma$. \cite{Wilting2018b,Zierenberg2018,Wilting2019,Das2019} Although a different criticality parameter will mostly be  used in the following, references to $m$ will continue for convenience.  

\begin{figure}
\includegraphics[width=\columnwidth]{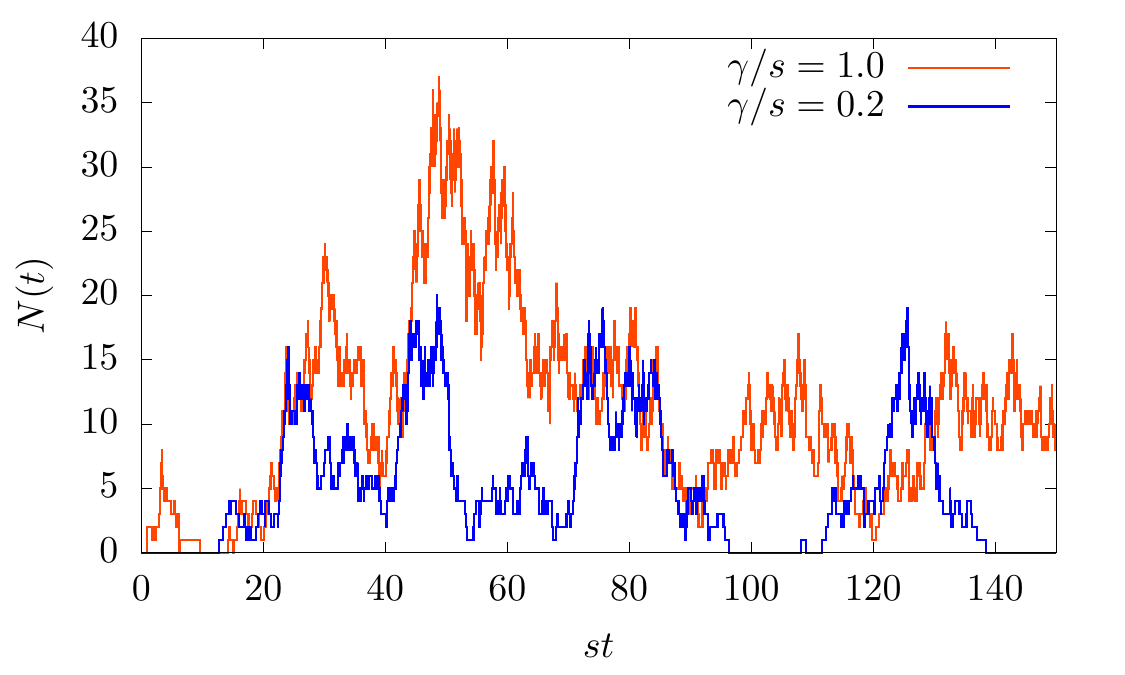}
\caption{Two example trajectories of a branching process with spontaneous creation. In both cases, the trajectories start with zero particles and have a binary offspring distribution with $r/s=0.1$. The spontaneous creation rate is $\gamma/s=1.0$ (red) and $\gamma/s=0.2$ (blue).}\label{fig-example-trajectory}
\end{figure}

The rate with which branching or extinction events of a single particle occur is denoted by $s$, i.e. the single particle extinction rate is $sp_0$ and the rate of creating two particles out of one is $sp_2$.\cite{Garcia-Millan2018} The difference between a branching process and a  branching process with immigration is that the latter includes spontaneous creation. The pumped branching process is simply the continuous-time version of the branching process with immigration. Example trajectories from Monte Carlo simulations of the pumped branching process are shown in Fig.~\ref{fig-example-trajectory}. Here, it can be seen that pumped branching does not have a fixed time bin size. The bin size is following an exponential distribution with rate $sp_0$, which allows implementing branching (with immigration) into continuous time. The interpretation of this model in terms of neuronal spike trains is that a spike occurs whenever an additional particle is created. Every inter-spike interval (ISI) has therefore a different realized time. Furthermore, the continuous-time implementation naturally leads to clustering of spikes, as the waiting time between branching events is shorter when there are already more particles in the system. In particular, this approach is not equivalent to adaptive binning \cite{Yu2017,Fontenele2019}, as such approaches still use fixed bin sizes within intervals.

How do different offspring distributions $p_k$ influence the phenomenology of the model? At a branching event, $k$ particles are created with probability $p_k$, replacing the original branching particle, that is $k-1$ additional particles are created. In particular, if the branching event creates two particles with probability $p_2$, then effectively \textit{only one additional} particle is created because the original branching particle is \textit{replaced} by two particles. The reason why only additional particles are important is that they are bosonic, i.e.~indistinguishable by definition. Therefore, the creation of $k$ particles at a branching event and the effective creation of $k-1$ additional particles must not lead to a different conclusion. This indistinguishability of particles manifests itself in the property that any choice of $p_1$ leads to the same stochastic process. If we include non-binary branching $p_k\neq0$ for $k\ge3$, then more than one additional particle can be created at the same branching event, i.e. at \textit{exactly} the same moment in time. However, signal propagation in the brain occurs on a continuous time spectrum across physical connections with a continuous length distribution. This length and time continuity implies that the inter-spike distribution should be continuous even when measured with a regular electrode grid. It is therefore fair to assume that signal creations in neurons never occur \textit{exactly} at the same time, and hence, it is reasonable to focus on binary branching only, i.e.~on the offspring distribution $p_j=0$ for all $j\notin\{0,2\}$.

As mentioned above, the major difference of the introduced model to other commonly used models is the continuous time distribution of events in place of fixed time steps. But how does this model compare to other continuous time models that are used in the analysis of neuronal spike recordings? The most important continuous time model in this context is the contact process and its variations \cite{Harris1974,Munoz2017,Nicoletti2020}. Contact processes describe how objects spread in discrete space, which usually is a regular lattice but which can be networks as well.  Thus, they lend themselves naturally as models of signal propagation in the brain. Gaining analytical results into the full model is extremely difficult and significant simplifications have to be made to understand its behaviour analytically. In the context of neuronal spikes, a contact process in a finite system with immigration has been used in \cite{Munoz2017}, while in \cite{Nicoletti2020} the contact process is without immigration and on an infinite 2D lattice. In contrast, the pumped branching process includes immigration but has no limit to its system size and has no geometry. In fact, a key component of contact processes in general and of the two mentioned studies is that an object that wants to spread only ever has a finite number of sites available to spread to.  A particle in a pumped branching process does not have these restrictions because there is no system size or system geometry. Thus, the pumped branching process could be regarded as the infinite dimensional limit of an infinite-size contact process with immigration.

In the remainder of this section, many important statistical properties of the pumped branching process are derived analytically using methods from statistical field theory. Although this novel derivation provides the necessary mathematical justification for the derivation of spike distributions and for the following data analysis, readers are invited to continue immediately with Sec.~\ref{sec-experiment} if they wish to skip the details of the field theory.


The resulting stochastic process of particle creation, branching and extinction can be described by a Doi-Peliti field theory.\cite{Doi1976,Peliti1985} Details on how to derive such field theories can be found in \cite{Taeuber2014,Garcia-Millan2018,Pausch2019b}. The branching and extinction part of the field theory have been described in \cite{Garcia-Millan2018}, where the resulting action $\mathcal{A}_\text{b}$ equals\begin{align}
\mathcal{A}_\text{b}=\int\widetilde\phi(t)\left(-\frac{\plaind}{\plaind t}-r\right)\phi(t)+\sum\limits_{j\ge2}q_j\widetilde\phi^j(t)\phi(t)\plaind t.\label{eq-branching-action}
\end{align}
Here, $\phi(t)$ is a particle annihilation field and $\widetilde\phi(t)$ is a Doi-shifted particle creation field. The effective extinction rate is $r=s(1-\mathbb{E}[K])$, which is zero at criticality. Since $s$ is a time scale which can be chosen arbitrarily, $r$ does not indicate how close the system is to criticality, but $\frac{r}{s}=1-m$ does. The system is subcritical for $\frac{r}{s}>0$ ($m<1$), supercritical for $\frac{r}{s}<0$ ($m>1$) and close to criticality if $\frac{r}{s}\approx0$ ($m\approx1$). Therefore, $\frac{r}{s}$ will be called \textit{degree of criticality}. The parameter $q_j=s\mathbb{E}[(K)_j]/j!$ is the $j$th factorial moment of the offspring distribution multiplyed by the branching rate $s$ and divided by $j!$. This article focuses on binary branching which implies that $q_j=0$ for all $j\ge3$.

The effects of spontaneous creation were not considered in \cite{Garcia-Millan2018} but can be added by including\begin{align}
\mathcal{A}_c=\int\gamma\widetilde\phi(t)\plaind t,\label{eq-spontaneous-creation-action}
\end{align}
where $\gamma$ is the spontaneous creation rate. In particular, it is the rate of an exponential distribution describing the times when external input (also called immigration or pumping) occurs. Its derivation is explained in Appendix~\ref{appendix-spontaneous-creation-derivation}. The dimensionless quantity $\frac{\gamma}{s}$ will be called \textit{relative spontaneous creation}. The action governing the dynamics of the entire system is $\mathcal{A}=\mathcal{A}_b+\mathcal{A}_c$. The path integral is normalized such that\begin{align}
\langle1\rangle=\int e^{\mathcal{A}t}\mathcal{D}[\widetilde\phi,\phi]=1,
\end{align}
and the expectation of an observable $\mathcal{O}$ is calculated as\begin{align}
\langle\mathcal{O}\rangle=\int \mathcal{O}e^{\mathcal{A}t}\mathcal{D}[\widetilde\phi,\phi].
\end{align}
In this field theory, the inclusion of spontaneous creation means that the system does not have to be initialized to show dynamics. Therefore, the easiest observables are $\mathcal{O}=\phi^m(t)$ which are the $m$th factorial moments of the particle number at time $t$. They are explained in Appendix~\ref{sec-factorial-moment-outline} and will play an important role in the following sections. 

The spontaneous creation leads to an average particle density $\zeta=\gamma/r$, and it is advantageous to shift the annihilation field $\phi=\check\phi+\zeta$ such that $\check\phi$ accounts for deviations from the average particle density. Then, the action does not contain a term proportional to $\widetilde\phi$ anymore, meaning that the shift of the annihilation field intrinsically  accounts for spontaneous creation.\cite{Pausch2019} The action $\mathcal{A}$ can be expressed in terms of this shifted field as\begin{align}
\mathcal{A}=\int \widetilde\phi(t)\left(-\frac{\plaind}{\plaind t}-r\right)\check\phi(t)+\sum\limits_{j\ge2}q_j\widetilde\phi^j(t)\left(\check\phi(t)+\zeta\right)\plaind t.\label{eq-final-action}
\end{align}
Eq.~\eqref{eq-final-action} shows the action on which all the following derivations are based. It represents a stochastic process of branching, extinction and spontaneous creation. The calculation of the expectations of observables can be represented by Feynman diagrams, which for Doi-Peliti field theories are read by convention from right to left. Given an action, the structure of occurring Feynman diagrams can be readily deduced by looking at the involved terms. Bilinear terms $\widetilde\phi(-\partial_t-r)\check\phi$ correspond to bare propagators and are represented by lines $\tikz[baseline=-2.5pt]{\draw[very thick,color=red] (-0.2,0.0) -- (0.1,0);}$ in the diagrams. All other terms are called interaction terms and are depicted as vertices, where the number of ingoing / outgoing  lines equals the power of $\check\phi$ / $\widetilde\phi$ in the interaction term. For example $q_2\widetilde\phi^2\check\phi$ is drawn as the vertex $\tikz[baseline=-2.5pt]{\draw[very thick,color=red] (-0.2,0.1) -- (0,0) -- (0.2,0);\draw[very thick,color=red] (-0.2,-0.1) -- (0,0);}$, and the vertex $\tikz[baseline=-2.5pt]{\draw[very thick,color=red] (-0.2,0.1) -- (0,0) -- (-0.2,-0.1);}$ represents the term $\zeta q_2\widetilde\phi^2$. Although the diagrams look like branching trees, it is important not to confuse a branching vertex with a branching event. The vertex represents correlations while the branching events are already incorporated in bare propagators. The vertices that are deduced from action $\mathcal{A}$ do not allow for diagrams with loops. Therefore, all observables that are polynomials in $\check\phi$ and $\widetilde\phi$ are represented by a finite number of diagrams. In particular, this implies that the bare propagator equals the full propagator.

What are some of the drawbacks of this model set-up? The model assumes stationary parameters. However, there are well known oscillations in the living brain \cite{Buzsaki2004}. Can it be expected that they alter the results of this analysis? The predominant oscillations are in the range between Delta and Gamma waves which cover a range of 0.1Hz to 100Hz. In Sec.~\ref{sec-experiment} experimental data from \cite{Buzsaki2009} is used to identify parameters. In this collection of data sets, the shortest recording is 17min long, while the longest is 106min. Therefore, it can be expected that effects from oscillations average out. In principle, this type of field-theoretic model can be adapted to accommodate time-dependent parameters, however, it comes at a cost of higher analytical difficulty.\cite{Pausch2020} Oscillations in the brain have been linked to the quiet times between avalanches and inter-avalanche correlations and repeated switches between up and down states\cite{Lombardi2012,Lombardi2014}, which do not appear in the pumped branching model because quiet times are exponentially distributed and therefore memoryless. This also implies that in this model, there are no inter-avalanche correlations.

The presented model is in a zero dimensional space. Could spatial components play a role? The answer is probably yes. In \cite{Klimm2014} it was found that embedded networks resemble the brain on a variety of network properties much more than non-embedded ones, in particular it was estimated that the topological fractal dimension of the human brain is 3.7 with high modularity and large global clustering. Another successful model is neutral theory, which includes a competition of causal avalanches for available sites \cite{Munoz-Plenz2017}. Such dynamics are not included in the present model. Furthermore, the pumped branching model does not contain any bounds on the number of active neurons. However, all experimental measurements are limited by the number of neurons in the capture area of the electrode array, resulting in finite size effects. Recently, branching processes with immigration (which are the discrete time versions of pumped branching processes) on networks have been compared to their zero-dimensional, macroscopic and simplified counterparts.\cite{Zierenberg2020} This study showed that macroscopic and microscopic parameters can differ significantly. However, it also showed in which parameter regions microscopic parameters of network dynamics are well approximated by the parameters of the zero-dimensional models. As the pumped branching model is closest to the `estimated rate' approximation in \cite{Zierenberg2020}, and since it will be shown in Sec.~\ref{sec-moment-ratio-map} that the data is in the subcritical regime, the bias introduced by ignoring network dynamics and coalescence effects is very small even with large spontaneous creation (see Fig.~(4) in \cite{Zierenberg2020}).

The dynamics can be divided into steady state and relaxation behavior. When the model is compared to experimental data in Sec.~\ref{sec-experiment}, it is assumed that data collection took place while the relevant brain region was in a steady state. Therefore the focus of the presented analytical derivations will be on steady state behavior. Relaxation behavior is still relevant though because it includes auto-correlations which have been used in  a new method for analyzing criticality in the brain, see Sec.~\ref{sec-twotimecorrelation}. Furthermore, the relaxation behavior is important when analytical results of the steady state are verified with Monte Carlo simulations, as is done throughout, because it indicates how long these simulations have to run to reach the steady state. 

\subsection{Steady state}
\label{sec-steady-state}
A steady state of a dynamic system is reached when the system is allowed to evolve for an infinite amount of time, $t\rightarrow\infty$. In general, such a state might not exist or it might depend on initial conditions. The following subsections show that the pumped branching process does have a steady state if the branching process is subcritical, i.e. $r>0$, which is assumed throughout.

\subsubsection{Moments}\label{sec-steadystatemoments}
In the system, particles are spontaneously created all the time with rate $\gamma$. Those particles undergo an effective extinction with rate $r$ resulting in a steady state average number of particles\begin{align}\label{eq-firstmoment}
\mathbb{E}[N]=\langle\phi\rangle=\underbrace{\langle\check\phi\rangle}_{=0}+\zeta=\frac{\gamma}{r},
\end{align}
which is expected in a system where effective extinction and spontaneous creation balance each other. 

This result shows a fundamental difference between a branching and a pumped branching process. Starting with a single particle in the system at $t=0$, in the infinite-time limit $t\rightarrow\infty$, the branching process has a discrete set of limit values for the mean particle number: $\lim_{t\rightarrow\infty}\mathbb{E}[N(t)|N(0)=1]\in\{0,1,\infty\}$ depending on $r>0$, $r=0$ or $r<0$. In contrast, the pumped branching process has a continuous set of limit values with $\lim_{t\rightarrow\infty}\mathbb{E}[N(t)|N(0)=1]=\frac{\gamma}{r}\sim r^{-1}$ as $r\rightarrow0^+$. In particular, the branching process has an infinite time survival probability of 0 for $r\ge0$, whereas the pumped branching process has a survival probability of~1. This shows, that the only steady state of the branching process is the absorbing state $N=0$, while the steady state of pumped branching process is dynamic. 

Calculating higher moments requires understanding the connection between factorial moments and moments.
The term $\langle\phi^k\rangle$ equals the $k$th factorial moment $\mathbb{E}[(N)_k]$ of the distribution, where $(x)_k=x(x-1)\cdots x(x-k+1)$ is the falling factorial. They are linked to the moments via the Stirling number of the second kind, $\begin{Bmatrix}n\\k\end{Bmatrix}$:\begin{align}\label{eq-fact-moment-sterling}
\mathbb{E}[N^n]=\sum\limits_{k=0}^n\begin{Bmatrix}n\\k\end{Bmatrix}\langle\phi^k\rangle.
\end{align}

Hence, the second moment equals\begin{align}\label{eq-secondmoment}
\mathbb{E}[N^2]=&\left\langle\phi^2\right\rangle+\left\langle\phi\right\rangle\,=\langle\check\phi^2\rangle+\zeta^2+\zeta\\
=&\,\frac{\gamma q_2}{r^2}+\frac{\gamma^2}{r^2}+\frac{\gamma}{r},
\end{align}
where the term $\langle\check\phi^2\rangle$ was calculated as follows\begin{align}
\left\langle\check\phi^2(t)\right\rangle\,\hat=&\,\,\tikz[baseline=-2.5pt]{\draw[very thick,color=red] (-1,0.4) -- (0,0) -- (-1,-0.4);}\\
=&\,2\zeta q_2\int\frac{\deltabar(\omega_1+\omega_2)e^{-(\omega_1+\omega_2)t}}{(-i\omega_1+r)(-i\omega_2+r)}\dbar\omega_1\dbar\omega_2\\
=&\,\frac{\gamma q_2}{r^2}.
\end{align}

\begin{figure}
\begin{center}
\includegraphics[width=\columnwidth]{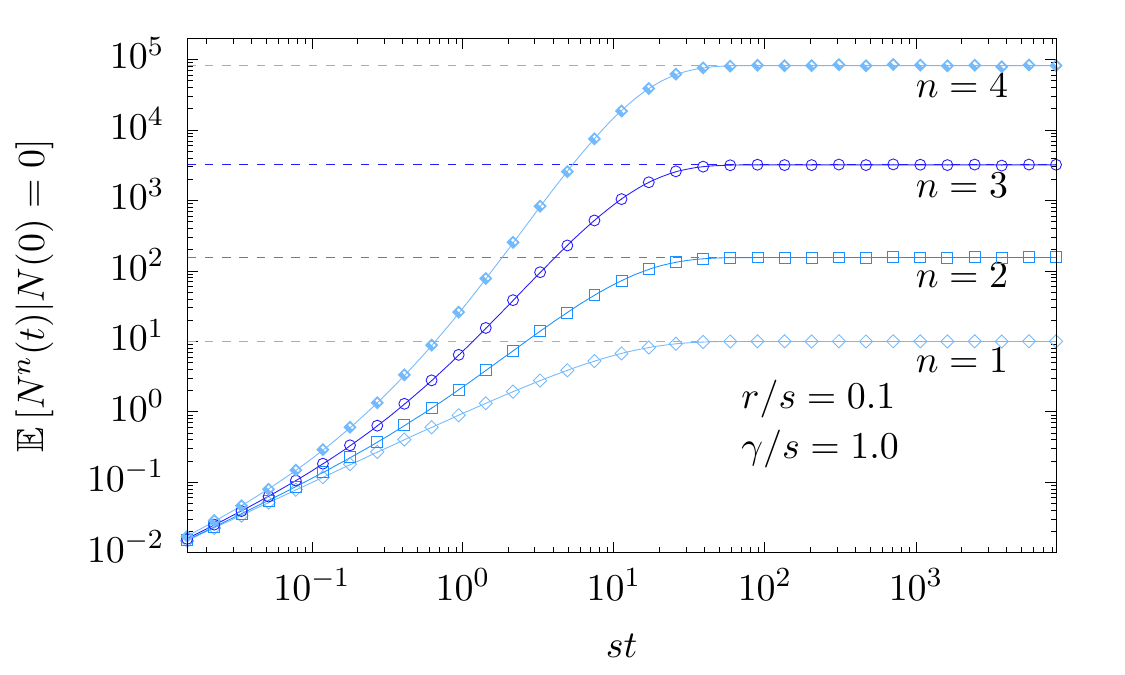}
\end{center}
\caption{1st, 2nd, 3rd and 4th moment of the process with binary offspring distribution with $r=0.1$ and $\gamma=1.0$. Dashed lines: steady state solutions. Solid lines: Relaxation starting with an empty system at $t=0$. Symbols: simulation results.}\label{fig-moments}
\end{figure}

Hence, the variance of the particle number is \begin{align}
\mathbb{V}ar\left(N(t)\right)=\frac{\gamma q_2}{r^2}+\frac{\gamma}{r}.
\end{align}
The first and second moment of the steady state are depicted as dashed lines in Fig.~\ref{fig-moments} with labels $n=1$ and $n=2$ respectively, alongside the 3rd and 4th moment which are derived below.  

Eq.~\eqref{eq-fact-moment-sterling} shows that calculating the $n$th moment requires calculating all the factorial moments up to the $n$th, which in turn require calculating all the $\langle\check\phi^\ell\rangle$ for $\ell\in\{1,\dots,n\}$. Their general expression can be found by exploiting a connection to cumulants.

\subsubsection{Factorial moments and factorial cumulants}\label{sec-cumulants}

Starting with the fourth moment, terms will appear which are represented by disconnected Feynman diagrams. In the fourth moment, the term $\langle\check\phi^4\rangle$ appears, which consists of the following Feynamn diagrams (if $q_j=0$ for $j\ge3$)\begin{align}\label{eq-disconnecteddiagrams}
\langle\check\phi^4\rangle\,\hat=\quad\tikz[baseline=-2.5pt]{\draw[very thick, color=red] (-0.5,0.2) -- (0,0.4) -- (-0.5,0.6); \draw[very thick, color=red] (-0.5,-0.6) -- (0,-0.4) -- (-0.5,-0.2);}\,\,+\,\tikz[baseline=-2.5pt]{\draw[very thick, color=red] (-0.5,0.2) -- (0,0.4) -- (-0.5,0.6); \draw[very thick, color=red] (-0.5,-0.6) -- (0,-0.4) -- (-0.5,-0.2); \draw[very thick, color=red] (0.0,0.4) -- (0.6,0) -- (0.0,-0.4);}\,+\,\tikz[baseline=-2.5pt]{\draw[very thick, color=red] (-0.5,0.2) -- (0,0.4) -- (-0.5,0.6); \draw[very thick, color=red] (0,0.4) -- (0.5,0.2) -- (-0.5,-0.2); \draw[very thick, color=red] (-0.5,-0.6) -- (1.0,0) -- (0.5,0.2);},
\end{align}
of which the first diagram consists of two disconnected diagrams. In order to find the general expression for $\langle\check\phi^n\rangle$, $h_n$ is defined as the part of $\langle\check\phi^n\rangle$ which is represented by connected Feynman diagrams only. Note that this can still include several diagrams, but they all start from a single $\zeta q_2\widetilde\phi^2\,\hat=\,\tikz[baseline=-2.5pt]{\draw[very thick,color=red] (-0.2,0.1) -- (0,0) -- (-0.2,-0.1);}$ vertex and end in $n$ outgoing lines. The diagrams representing $h_n$ differ only by their internal diagrammatic structure, see for example the 2nd and 3rd diagram in Eq.~\eqref{eq-disconnecteddiagrams}. Using \tikz[baseline=-2.5pt]{\draw[very thick, color=red] (-0.3,0.1) -- (0,0) -- (0.3,0);\draw[very thick, color=red] (-0.3,-0.1) -- (0,0);\filldraw[fill=blue] (0,0) circle[radius=0.1cm];} to hide all the possible internal diagrammatic structure, $h_n$ is is written as\begin{align}
h_n\,\hat=&\,\sum\limits_{m=1}^{n-1}{n\choose m}\tikz[baseline=-2.5pt]{\draw[very thick, color=red] (-1,0.4) -- (0,0) -- (-1,-0.4);\draw[very thick, color=red] (-2,0.7) -- (-1,0.4) -- (-2,0.1); \draw (-2.0,0.5) node {$\vdots$};\draw (-2,0.4) node[left] {$m$};\draw[very thick, color=red] (-2,-0.7) -- (-1,-0.4) -- (-2,-0.1); \draw (-2.0,-0.3) node {$\vdots$};\draw (-2,-0.4) node[left] {$n-m$};\filldraw[fill=blue] (-1,0.4) circle[radius=0.2cm];\filldraw[fill=blue] (-1,-0.4) circle[radius=0.2cm];}\\
=&\sum\limits_{m=1}^{n-1}{n\choose m}\zeta q_2\int\limits \tilde g_m(\omega)\tilde g_{n-m}(-\omega)\dbar\omega\\
=&\sum\limits_{m=1}^{n-1}{n\choose m}\zeta q_2\int\limits g_m(t)g_{n-m}(t)\plaind t,
\end{align}
where $\tilde g_m(\omega)$ is the propagator of the branch with one ingoing leg and $m$ outgoing legs in Fourier space:\begin{align}
\tilde g_m(\omega)\hat=\tikz[baseline=0.33cm]{\draw[very thick, color=red] (-2,0.7) -- (-1,0.4) -- (-2,0.1); \draw[very thick, color=red] (-1,0.4) -- (0,0.4); \draw (-2.0,0.5) node {$\vdots$};\draw (-2,0.4) node[left] {$m$};\filldraw[fill=blue] (-1,0.4) circle[radius=0.2cm];}.
\end{align}  
Its real space version $g_m(t)$ was found in \cite{Garcia-Millan2018} to be \begin{align}\label{eq-internal-factorial-moment}
g_m(t)=m!e^{-rt}\left(\frac{q_2}{r}\left(1-e^{-rt}\right)\right)^{m-1}.
\end{align}
Hence, $h_n$ equals \begin{align}\label{eq-connecteddiagramterm}
h_n=\zeta(n-1)!\left(\frac{q2}{r}\right)^{n-1},
\end{align} 
which is true for $n>1$. For $n=1$, $\langle\check\phi\rangle=0$, but we set $h_1=\zeta=\frac{\gamma}{r}$. This choice means that the $\zeta$ in the shift $\langle\phi^n\rangle=\langle(\zeta+\check\phi)^n\rangle$ is interpreted as its own subdiagram, which turns out to be useful in the following. If non-binary offspring distributions were chosen, then the internal diagrammatic structure would be more complicated and involve $q_3,q_4,\dots$. The corrections to $g_m$ in Eq.~\eqref{eq-internal-factorial-moment} would be of the order $\mathcal{O}(1/r^{m-2})$, which implies that $g_m$ is still a good approximation close to criticality $r\approx0^+$. This means that $h_n$ is also a good approximation for non-binary offspring distributions for $r\approx0^+$  and its error is $\mathcal{O}(1/r^{n-1})$.

In the following, the terms $h_n$, which are represented by the connected diagrams, are shown to be  \textit{factorial}  cumulants. This argument is used in an analogous way for the relation between moments and cumulants in \cite{LeBellac1991}.

First, let's define the factorial moment partition function:\begin{align}
Z_\phi(j)=\sum\limits_{n=0}^\infty\frac{j^n\langle\phi^n\rangle}{n!},
\end{align}
which is \textit{different} to the standard definition of the factorial moment generating function in that the factorial moments are its derivatives at the origin.
Analogously, the factorial cumulant generating function is defined as $W_\phi(j)=\ln Z_\phi(j)$, and thus factorial cumulants $\langle\phi^k\rangle_c$ are also defined:\begin{align}
W_\phi(j)=\sum\limits_{n=0}^\infty\frac{j^n\langle\phi^n\rangle_c}{n!}.
\end{align}

Let's assume, we want to calculate $\langle\phi^n\rangle$. Then, we have to sum over many diagrams, all of which have $n$ outgoing legs, and some of which won't be connected diagrams. They will consist of a few connected subdiagrams. Let's group these connected subdiagrams such that the $j$th subdiagram has $n_j$ outgoing legs, and the diagram is repeated $q_j$ times in the big diagram. Hence $n=q_1n_1+\dots q_pn_p$, where $p$ is the number of types of connected subdiagrams. The types of subdiagrams are only determined by the number of outgoing legs. Depending on the action, there might be several types of connected diagrams with the same number of outgoing legs but with differing internal structure. For the purpose of the following calculation, they are grouped as one diagram and their combined Green's function $G_c^{(n_j)}$ is the sum of the Green's function of each diagram within the group.

The number of independent terms is $n!$ (the number of permutations of all legs) divided by the number of permutations of the legs within each subdiagram, i.e. $n_j!$, and the number of permutations within the same type of subdiagram, denoted by $q_j!$:\begin{align}
\text{\#  independent terms }=\frac{n!}{(n_1!)^{q_1}q_1!\dots (n_p!)^{q_p}q_p!}
\end{align}
For now, we allow the times at which the outgoing legs are evaluated to be arbitrary, i.e the $t_1,t_2,\dots$ in $\langle\phi(t_1)\dots\phi(t_n)\rangle$ don't have to be equal. Then the partition function for the factorial moments is\begin{align}\begin{split}
&Z_\phi(j)=\sum\limits_{n=0}^\infty\frac{1}{n!}\int\plaind^n\mathbf{t} j(t_1)\dots j(t_n)\times\\
&\hspace{-0.3cm}\sum\limits_{\sum_{\ell=1}^pq_\ell n_\ell=n}\hspace{-0.65cm}G_c^{(n_1)}(t_1,\dots,t_{n_1})\dots G_c^{(n_p)}(t_{n-n_p+1},\dots,t_n),
\end{split}\end{align}
where $G_c^{(n_j)}$ is the Green's function of the connected subdiagrams of type $j$, which have $n_j$ outgoing legs, and where $\mathbf{t}=(t_1,\dots,t_n)$. This expression can be rearranged such that\begin{align}
Z_\phi(j)=\exp\Biggl(\sum\limits_{n=1}^\infty\frac{1}{n!}\int j(t_1)\dots j(t_n)G_c^{(n)}(\mathbf{t})\plaind^n\mathbf{t}\Biggr).
\end{align}
Hence, the factorial cumulant generating function is identified as\begin{align}
W_\phi(j)=\sum\limits_{n=1}^\infty\frac{1}{n!}\int j(t_1)\dots j(t_n)G_c^{(n)}(\mathbf{t})\plaind^n\mathbf{t},
\end{align}  
with  $G_c^{(n)}(t_1,\dots,t_n)=\langle\check\phi(t_1)\dots\check\phi(t_n)\rangle_c$.

The terms corresponding to the connected Feynman diagrams were found in Eq.~\eqref{eq-connecteddiagramterm}. Therefore, the generating functions $W_\phi$ and $Z_\phi$ can be written as\begin{align}
W_\phi(j)=&\sum\limits_{n=1}^\infty\frac{j^nh_n}{n!}=\sum\limits_{n=1}^\infty\frac{j^n\gamma q_2^{n-1}}{n r^n}\\
Z_\phi(j)=&\exp\left({\sum\limits_{n=1}^\infty\frac{j^n\gamma q_2^{n-1}}{nr^n}}\right),
\end{align}
which allows finding the $k$th factorial moment by taking the $k$th derivative w.r.t $j$ and evaluating at $j=0$:\begin{align}
\langle\phi^k\rangle=&\,\frac{\plaind^k}{\plaind j^k}\exp\left({\sum\limits_{n=1}^\infty\frac{j^n\gamma q_2^{n-1}}{nr^n}}\right)\Biggr|_{j=0}\\
=&\,\left(\frac{q_2}{r}\right)^k\sum\limits_{\ell=0}^k\begin{bmatrix}k\\\ell\end{bmatrix}\left(\frac{\gamma}{q_2}\right)^\ell\\
=&\,\left(\frac{q_2}{r}\right)^k\left(\frac{\gamma}{q_2}\right)^{(k)},
\end{align}
where $\begin{bmatrix}k\\\ell\end{bmatrix}$ is the unsigned Stirling number of the first kind and $x^{(k)}=x(x+1)\cdots(x+k-1)$ is the rising factorial. The first four factorial moments are\begin{align}
\langle\phi\rangle=&\,\frac{\gamma}{r}\\
\langle\phi^2\rangle=&\,\frac{\gamma^2}{r^2}+\frac{\gamma q_2}{r^2}\\
\langle\phi^3\rangle=&\,\frac{\gamma^3}{r^3}+\frac{2\gamma q_2^2}{r^3}+\frac{3\gamma^2 q_2}{r^3}\\
\langle\phi^4\rangle=&\,\frac{6\gamma q_2^3}{r^4}+\frac{11\gamma^2 q_2^2}{r^4}+\frac{6\gamma^3 q_2}{r^4}+\frac{\gamma^4}{r^4},
\end{align}
confirming the previous calculation of the first and second moment, Eqs.~\eqref{eq-firstmoment} and \eqref{eq-secondmoment}. The implied 3rd and 4th moment are also shown in Fig.~\ref{fig-moments}.

Note that with a binary offspring distribution ($q_j=0\forall j\ge3$), the $k$th factorial moment diverges like $r^{-k}$ as $r\rightarrow0$. For non-binary offspring distributions ($\exists j\ge3$ with $q_j\neq0$), the error is $\mathcal{O}(1/r^{k-1})$. This implies that at criticality $r=0$, there are universal factorial moment ratios which are also universal moment ratios due to their connection shown in Eq.~\eqref{eq-fact-moment-sterling}, for example:\begin{align}
\lim\limits_{r\rightarrow0^+}\frac{\mathbb{E}[N^n]}{\mathbb{E}[N^m]\mathbb{E}[N^{n-m}]}=&\,\lim\limits_{r\rightarrow0^+}\frac{\langle\phi^n\rangle}{\langle\phi^m\rangle\langle\phi^{n-m}\rangle}\\=&\,\frac{B\left(\frac{\gamma}{q_2},m\right)}{B\left(\frac{\gamma}{q_2}+n-m,m\right)},
\end{align}
with the Beta function $B(x,y)=\Gamma(x)\Gamma(y)/\Gamma(x+y)$.

\subsubsection{Probability to be in state $N=n$}\label{sec-steadystateprobabilitydistribution}
In this section, the steady state probability distribution $P(N=n)$ for the system to be in state $N=n$  is derived. The calculation is based on the probability generating function $\mathcal{M}(z)$\begin{align}
\mathcal{M}(z)=\sum\limits_{\ell=0}^\infty P(N=\ell)z^\ell.
\end{align}
It does not only encode the probability distribution $P(N=n)$, but also the factorial moments:\begin{align}
\frac{\plaind^k}{\plaind z^k}\mathcal{M}(z)\Bigr|_{z=1}\hspace{-0.26cm}=\sum\limits_{\ell=k}^\infty \frac{\ell !}{k!}P(N=\ell)=\mathbb{E}[(N)_k]=\langle\phi^k\rangle.
\end{align}
As $\mathcal{M}(z)$ is analytic, $P(N=n)$ can be derived from the factorial moments:\begin{align}
P(N=\ell)=&\frac{1}{\ell!}\frac{\plaind^\ell}{\plaind z^\ell}\sum\limits_{k=0}^\infty\frac{(z-1)^k}{k!}\langle\phi^k\rangle\label{eq-steadystateprobability1}\\
=&\frac{1}{\ell!}\left(\frac{q_2}{r}\right)^\ell\left(\frac{\gamma}{q_2}\right)^{(\ell)}\left(1+\frac{q_2}{r}\right)^{-\frac{\gamma}{q_2}-\ell}\label{eq-steadystateprobability2}\\
=&\frac{\Gamma\left(\frac{\gamma}{q_2}+\ell\right)}{\ell!\Gamma\left(\frac{\gamma}{q_2}\right)}\left(\frac{r}{r+q_2}\right)^{\frac{\gamma}{q_2}}\left(\frac{q_2}{r+q_2}\right)^\ell. \label{eq-steadystateprobability3}
\end{align}
The step from Eq.~\eqref{eq-steadystateprobability1} to Eq.~\eqref{eq-steadystateprobability2}, and the check that it is normalized is shown in Appendix~\ref{appendix-steadystateprobability}. This probability distribution is exact for the process with binary offspring distribution and satisfies its steady-state master equation, Eq.~\eqref{eq-appendix-master-equation} and~\eqref{eq-master-eq-branching}. It is a negative Binomial (or P\'olya) distribution with clustering coefficient $\frac{\gamma}{q_2}$. Small cluster coefficients imply common occurrence of bursts of activity, while large cluster coefficients imply activity dispersion. 

For $q_2\rightarrow0$, the branching process is turned off and the pure Poisson process is found. In particular, $P(N=\ell)$ becomes a Poisson distribution \begin{align}\lim\limits_{q_2\rightarrow0}P(N=\ell)=\frac{1}{\ell!}\left(\frac{\gamma}{s}\right)^\ell e^{-\frac{\gamma}{s}}.
\end{align}
For $r\rightarrow0^+$, $P(N=\ell)\sim r^{\frac{\gamma}{q_2}}\rightarrow 0$ because the number of particles in the system diverges.

\begin{figure}
\begin{center}
\includegraphics[width=\columnwidth]{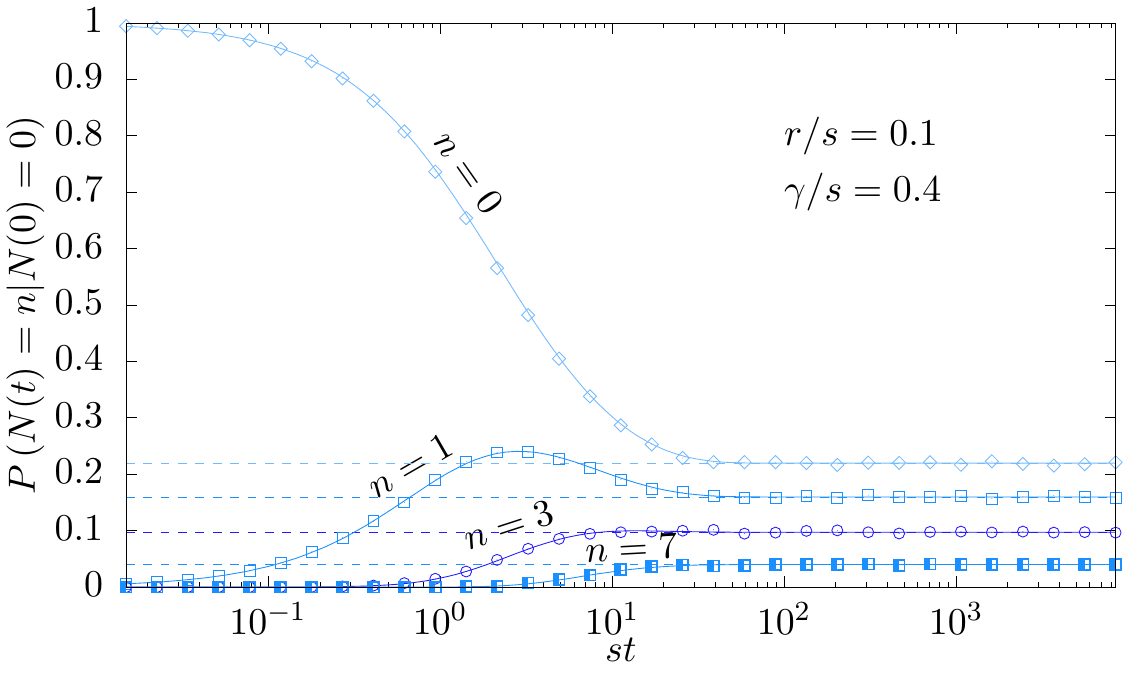}
\end{center}
\caption{Probability that there are $n=0,1,3,7$ particles in a system using a binary offspring distribution with $\frac{r}{s}=0.1$ and $\frac{\gamma}{s}=0.4$. Dashed lines: steady state solutions. Straight lines: Relaxation starting with an empty system at $t=0$. Symbols: simulation results.}\label{fig-probabilities}
\end{figure}

Examples of the steady-state probabilities are shown in Fig.~\ref{fig-probabilities} and Fig.~\ref{fig-frequency}.

\subsubsection{Expected avalanche duration and size}\label{sec-avalanche-length-and-size}
Avalanches in the brain are one of the most studied observables.\cite{Beggs2003,Beggs2004,Beggs2008,Priesemann2009,Priesemann2013,Lombardi2014} However, retrieving statistical information about them that is stable under small variations of the processing parameters has been challenging.\cite{Priesemann2009} Because of their prominence, some analytical insight is likely to be useful. 

\begin{figure}
\begin{center}
\includegraphics[width=\columnwidth]{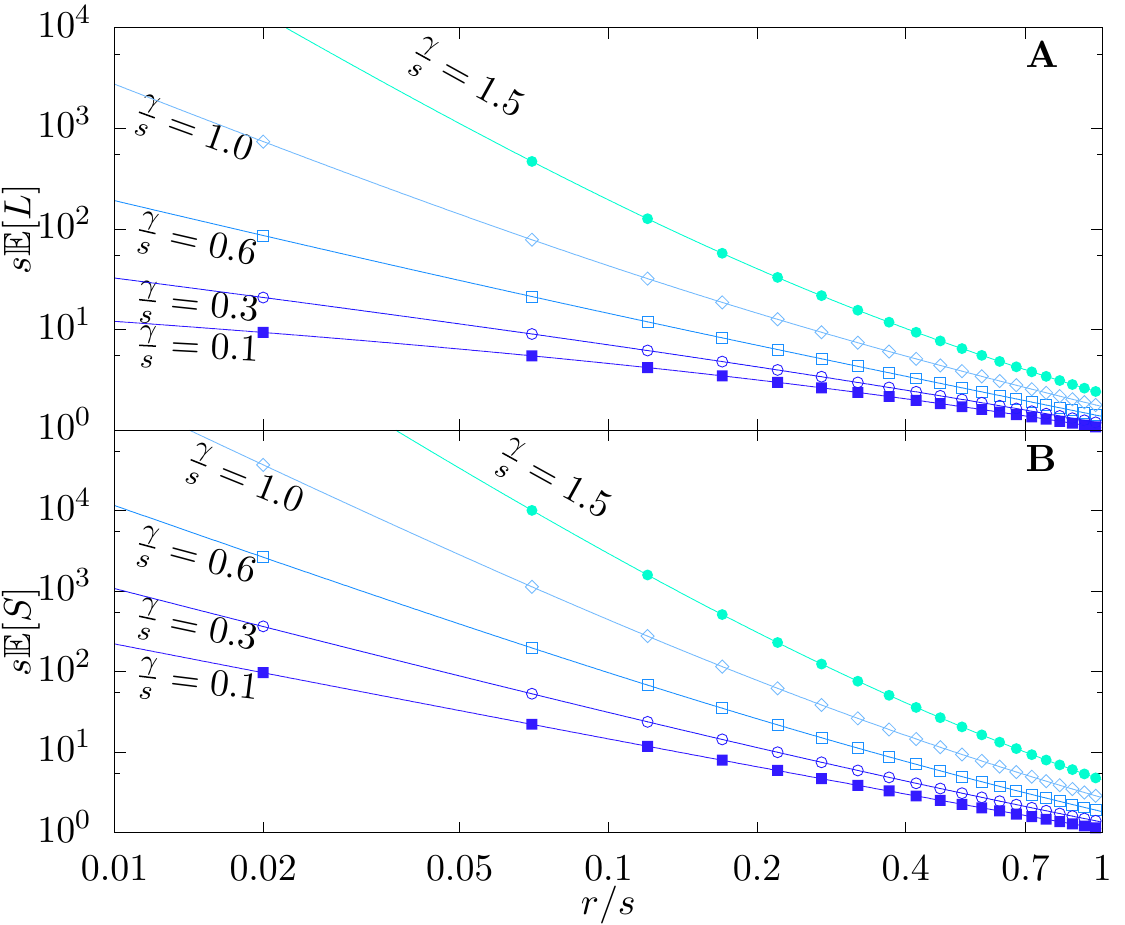}
\end{center}
\caption{Expected avalanche duration $L$ (Panel \textbf{A}) and size $S$ (Panel \textbf{B}).  Using a binary offspring distribution, the chosen parameters are $\frac{\gamma}{s}\in\{0.1,0.3,0.6,1.0,1.5\}$. Straight lines: analytical results. Symbols: simulation results.}\label{fig-lengths}
\end{figure}

Avalanches are characterized by their duration $L$ and size $S$. The expected duration of an avalanche is defined as the time between leaving state $N=0$ and entering it again. It can be derived from the probability distribution $P(N=n)$ by considering that the ratio of the time spent in state $N\neq0$, i.e. in an avalanche, to the time spent in $N=0$, i.e. in between avalanches, is equal to the ratio of the probabilities of these states:\begin{align}
\frac{\mathbb{E}[\text{time spent in avalanches}]}{\mathbb{E}[\text{time spent between avalanches}]}=\frac{P(N\neq0)}{P(N=0)}.
\end{align}
Since the number of avalanche gaps is equal to the number of avalanches, the time spent in a gap per avalanche is the expected time of a spontaneous creation to occur $1/\gamma$. Therefore the expected avalanche duration equals \begin{align}\label{eq-expected-length}
\mathbb{E}[L]=\frac{1}{\gamma}\frac{P(N\neq0)}{P(N=0)}=\frac{1}{\gamma}\left(\left(1+\frac{q_2}{r}\right)^{\frac{\gamma}{q_2}}-1\right).
\end{align}
In the limit $\gamma\rightarrow0$, where gaps between avalanches become infinitely long, the statistics of a pure branching process is recovered $\lim_{\gamma\rightarrow0}\mathbb{E}[L]=\ln(1+q_2/r)/q_2$.\cite{Garcia-Millan2018} In the limit where branching ceases to occur, i.e. $p_0\rightarrow1$ ($r/s\rightarrow1$, $q_2/s\rightarrow0$), the pure Poisson process is found $\lim_{p_0\rightarrow1}\mathbb{E}[L]=(e^{\frac{\gamma}{s}}-1)/\gamma$. Within this Poisson limit, the case $\gamma\rightarrow0$ describes a system without branching where spontaneous creations never happen. In this limit, the avalanche duration is the expected extinction time of a single particle $\lim_{\gamma\rightarrow0}\lim_{p_0\rightarrow1}\mathbb{E}[L]=\frac{1}{s}$. The expected avalanche duration is shown in Fig.~\ref{fig-lengths} along side simulations.

Eq.~\eqref{eq-expected-length} also implies that avalanches can be considered as an overlap of several causal avalanches, where a causal avalanche is defined as a subset of particles in the system that originated from the same single particle through \textit{branching only}. This definition means that causal avalanches are initiated by a spontaneous creation and then follow the branching-extinction process. Every new spontaneous creation during an avalanche will start a new causal avalanche within it. Therefore, Eq.~\eqref{eq-expected-length} implies that an avalanche consists on average of $\gamma\mathbb{E}[L]$ causal avalanches.\cite{Munoz-Plenz2017}

The other avalanche property of interest is its size, i.e. the integral of the particle number over its entire duration. The expected size can be derived analogously to the duration by considering that an avalanche spends on average $P(N=n)/(\gamma P(N=0))$ amount of time in state $N=n$. Therefore its expected size is\begin{align}\label{eq-expected-size}
\mathbb{E}[S]=\sum\limits_{n=1}^\infty\frac{nP(N=n)}{\gamma P(N=0)}=\frac{1}{r}\left(1+\frac{q_2}{r}\right)^{\frac{\gamma}{q_2}}.
\end{align}
On the one hand, the limit $\gamma\rightarrow0$ describes the infinite separation of avalanches and recovers the expected avalanche size of the pure branching process.\cite{Garcia-Millan2018}  On the other hand, eliminating branching by taking the limit $p_0\rightarrow1$ ($r/s\rightarrow1$, $q_2/s\rightarrow0$) gives the pure Poisson process result $\lim_{p_0\rightarrow1}\mathbb{E}[S]=e^{\frac{\gamma}{s}}/s$.

\subsection{Relaxation towards steady state}\label{sec-relaxation}
In the subcritical regime $r>0$, the system has an active steady state which was characterized in Secs.~\ref{sec-steadystatemoments} and~\ref{sec-steadystateprobabilitydistribution}. In this section, it is calculated how the system converges to the steady state. Aside from theoretical interest, the relaxation behavior deserves attention for three practical reasons. 

First, the brain might switch between steady states, e.g. between sleep and wakefulness, and its signal propagation would take time to reach the new steady state. Second, analytical results in this article are verified by Monte Carlo simulations. Each simulation starts with a specific initial condition and needs to run for some time to be a good approximation of the steady state.  Third, one of the methods for determining the degree of criticality is based on auto-correlation functions \cite{Wilting2018b} and it is of scientific interest to check whether different approaches reach similar conclusions. 

Often dimensional analysis is sufficient to identify such time scales. However, the pumped branching process contains several time scales $r$, $q_j$ and $\gamma$, which can be vastly different. As the following subsections show, the relaxation time is solely determined by $r=s(1-\mathbb{E}[K])$, the other time scales $q_j$ and $\gamma$ do not enter. 

\subsubsection{Single particle insertion}
Let's assume that the system is in steady state. If a particle is placed into the system by hand, then the expected number of particles decays exponentially to the steady state particle number:\begin{align}\begin{split}
\mathbb{E}[N(t)|N(0^-)+1=N(0^+)]=\hspace{1cm}\\=\left\langle\phi(t)\phi^\dagger(0)\right\rangle=\zeta+\Theta(t)e^{-rt},
\end{split}\end{align}
where $\Theta(t)$ is the Heaviside function and the superscript `$-$' and `$+$' denote one-sided limits. Here $N(0)$ is the unknown number of particles which the steady-state system contains just before an additional particle is inserted into system. The insertion thus pushes it out of its steady-state dynamics.

\subsubsection{Two-Time auto-correlation}\label{sec-twotimecorrelation}
In the steady state, particles are created and go extinct at random. However, the process cannot jump arbitrarily between states. In particular with binary branching, the particle number $N$ can only increase or decrease by one particle at a time, which induces time-correlations within the steady state. The simplest of these is the two-time auto-correlation\begin{align}
\text{corr}(t_1,t_2)=\mathbb{E}[N(t_1)N(t_2)].
\end{align}
It can be calculated as\begin{align}\begin{split}
\text{corr}(t_1,t_2)=&\Theta(t_1-t_2)\langle\phi(t_1)\phi^\dagger(t_2)\phi(t_2)\rangle\\
&+\Theta(t_2-t_1)\langle\phi(t_2)\phi^\dagger(t_1)\phi(t_1)\rangle\end{split}\\
=&\frac{\gamma^2}{r^2}+\frac{\gamma q_2}{r^2}e^{-r|t_1-t_2|}+\frac{\gamma}{r} e^{-r|t_1-t_2|},\label{eq-2time-corr}
\end{align}
where the second part equals
\begin{align}
\langle\phi(t_2)\phi^\dagger(t_1)\phi(t_1)\rangle=\zeta^2+\langle\check\phi(t_2)\check\phi(t_1)\rangle+\zeta\langle\check\phi(t_2)\widetilde\phi(t_1)\rangle.
\end{align}

\begin{figure}
\begin{center}
\includegraphics[width=\columnwidth]{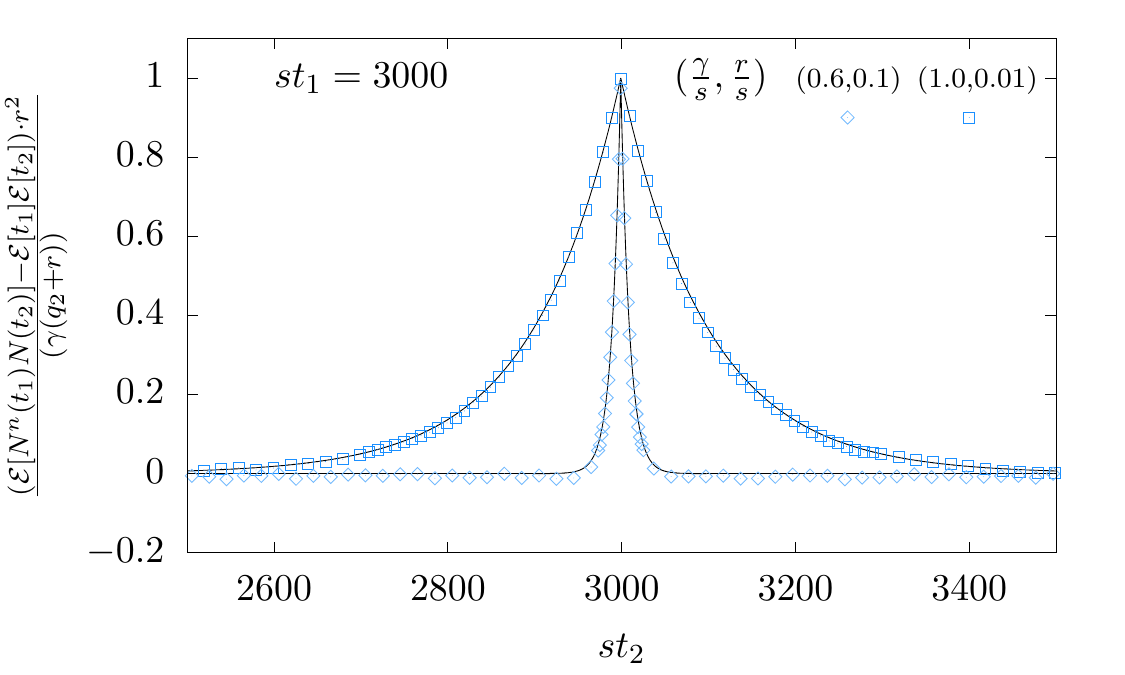}
\end{center}
\caption{The rescaled correlation function $\text{corr}(t_1,t_2)$ is shown for the binary offspring distributions with $r/s=10^{-1}$ and $\gamma/s=0.6$ as well as $r/s=10^{-2}$ and $\gamma/s=1.0$. Straight lines:  analytical prediction. Symbols: simulation results for trajectories which started with zero particles at time $t=0$. The time $st_1=3000$ was chosen in order to be well in the steady state regime.}\label{fig-twotimecorrelation}
\end{figure}

Two examples of  rescaled correlation functions are shown in Fig.~\ref{fig-twotimecorrelation}. 

The autocorrelation function, Eq.~\eqref{eq-2time-corr}, implies a specific relation between the intrinsic time scales $1/r$ of autocorrelations and its offset $\gamma^2/r^2$: assuming that spontaneous creation $\gamma$ stays constant or undergoes comparatively small changes, larger intrinsic time scales imply larger offsets. This analytic relation matches with the experimental observation in \cite{Murray2014} (specifically Fig.~2b in \cite{Murray2014}), where intrinsic time scales and offsets of autocorrelations where compared across several brain regions and a hierarchy was found.

What does this result, Eq.~\eqref{eq-2time-corr}, imply for the method of estimating the systems closeness to criticality in \cite{Wilting2018b}? In \cite{Wilting2018b}, the criticality of a pumped branching process (or branching process with immigration) in discrete time was estimated using the discrete time auto-correlation function. Using discrete time fixes the time scale of activity of the system. Choosing a time scale does make sense for systems with a natural / practical time scale, e.g. choosing days for the spread of infectious diseases. Whether there is such a time scale for signal propagation in neuronal networks is not clear, although many authors use the average inter-spike interval $\langle\text{ISI}\rangle$. 

A method that samples the continuous auto-correlation function in Eq.~\eqref{eq-2time-corr} in discrete time steps $\Delta t$, assumes the relation $e^{-rk\Delta t}=\widehat{\mathbb{E}[K]}^k$ and erroneously infers $\widehat{\mathbb{E}[K]}=e^{-r\Delta t}\approx1-r\Delta t$, where $\widehat{\hspace{0.2cm}}$ denotes an estimator. Thus, it estimates $\widehat{\mathbb{E}[K]}=s\Delta t \mathbb{E}[K]+1-s\Delta t$. Therefore, $\widehat{\mathbb{E}[K]}$ is only unbiased if the chosen time step is $\Delta t=\frac{1}{s}$. However, $s$ cannot be inferred separately from $r=s(1-\mathbb{E}[K])$ when the auto-correlation function Eq.~\eqref{eq-2time-corr} is used. In fact, in the supplementary notes of \cite{Wilting2018b}, it was noted that $\widehat{\mathbb{E}[K]}$ (called $m$ in the article) scales with $\Delta t$.

However, when branching-like processes are generated from spike time series \cite{Beggs2003,Beggs2004,Priesemann2009,Priesemann2013,Wilting2018a,Wilting2018b}, the bin size $\Delta t$  represents the time between creation (spike) and extinction of a particle, Fig.~\ref{fig-schematic}. In a continuous branching process, the average time between creation and extinction is $\frac{1}{sp_0}$, the inverse of the extinction frequency.   

Therefore, choosing $\Delta t=\frac{1}{sp_0}$ to create a branching-like process and using it to infer the expected offspring number $\mathbb{E}[K]$ from the auto-correlation function (i.e. assuming $\Delta t=\frac{1}{s}$) is contradicting itself except if it is a Poisson process with $p_0=1$. Most publications \cite{Beggs2003,Beggs2004,Priesemann2009,Priesemann2013,Wilting2018a,Wilting2018b} find the system to be close to or at criticality, which means that $p_0\approx\frac{1}{2}$. 

Furthermore, most publications assume $\Delta t=\langle\text{ISI}\rangle$ \cite{Beggs2003,Beggs2004,Priesemann2009,Priesemann2013,Lombardi2014,Wilting2018a,Wilting2018b} but it remains unclear whether this would be the right choice for generating a branching-like processes in the first place. This question will be addressed in Sec.~\ref{sec-ISI-extinction-time}.

In Sec.~\ref{sec-moment-ratio-map}, the estimation based on auto-correlation is applied to the pumped branching process and indeed a bias is found in the presented examples, see Figs.~\ref{fig_ComparisonContinuousDiscrete} and~\ref{fig_ComparisonContinuousDiscreteMultiStep}.

\subsubsection{Initiating an empty system}\label{sec-empty-system}
Due to the spontaneous creation with rate $\gamma$, the system is only occasionally empty and, at an arbitrarily chosen time in the steady state, it is empty with probability $P(N=0)<1$. However, an empty system can be enforced at time $t=0$ by inserting the factor $e^{-\widetilde\phi(0)\phi(0)}$ in front of observables, resulting in $P(N(0)=0)=1$. The derivation of this adjustment is outlined in Appendix~\ref{sec-appendix-empty-systen}. Given the system was empty at $t=0$, the $n$th factorial moment at time $t$ is\begin{align}\label{eq-factorialmomentsrelaxation}
\mathbb{E}[\left(N(t)\right)_n|N(0)=0]=\left\langle \phi^n(t)e^{-\widetilde\phi(0)\phi(0)}\right\rangle.
\end{align}
If $e^{-\widetilde\phi(0)\phi(0)}$ is written as a series, and if the allowed Feynman diagram structures are considered, it can be deduced that the $n$th factorial moment will only have non-zero contributions from the first $n+1$ terms of the series expansion, i.e. the terms $1-\widetilde\phi(0)\phi(0)+\dots+(-\widetilde\phi(0)\phi(0))^n/n!$. For this calculation it is advantageous to reverse the shift of the annihilation field $\check\phi=\phi-\zeta$ and use the original action from Eq.~\eqref{eq-branching-action} and~\eqref{eq-spontaneous-creation-action}. To highlight this difference the term $\gamma\widetilde\phi$, which does not appear in the $\check\phi$-shifted action, is represented by a \tikz[baseline=-2.5pt]{\draw[very thick, color=red](-0.2,0) -- (0,0) node{\color{blue}$\times$};} vertex.
For the 1st and 2nd factorial moment, the Feynman diagrams resulting from Eq.~\eqref{eq-factorialmomentsrelaxation} are shown in the following:\begin{align}
\left\langle\phi(t)\widetilde\phi(0)\phi(0)\right\rangle\,\hat=&\,\tikz[baseline=-2.5pt]{\draw[very thick,color=red] (-0.9,0) -- (-0.1,0); \draw[very thick,color=red] (0.0,0)  -- (0.5,0) node {\color{blue}$\times$};}\notag\\
=&\zeta e^{-rt}\\
\left\langle\phi^2(t)\widetilde\phi(0)\phi(0)\right\rangle\,\hat=&\,\tikz[baseline=-2.5pt]{\draw[very thick,color=red] (-0.6,0.3) -- (0,0) -- (0.5,0); \draw[very thick,color=red] (-0.6,-0.3) -- (0,0); \draw[very thick,color=red] (0.6,0) -- (1.1,0) node {\color{blue}$\times$};}+\tikz[baseline=-2.5pt]{\draw[very thick,color=red] (-0.7,0.3) -- (-0.2,0.3); \draw[very thick,color=red] (-0.7,-0.3) -- (-0.2,-0.3) -- (0.2,0.3); \draw[very thick,color=red] (-0.1,0.3) -- (0.6,0.3) node {\color{blue}$\times$}}+\tikz[baseline=-2.5pt]{\draw[very thick,color=red] (-0.5,0.3) -- (0,0.3); \draw[very thick,color=red] (0.1,0.3) -- (0.6,0.3) node {\color{blue}$\times$}; \draw[very thick,color=red] (-0.5,-0.3) -- (0.6,-0.3) node {\color{blue}$\times$};}\notag\\
&\hspace{-2.3cm}=2\zeta\frac{q_2}{r}e^{-rt}\left(1-e^{-rt}\right)+2\zeta\frac{q_2}{r}e^{-2rt}+\zeta^2e^{-rt},\label{eq-secondmomentrelaxation}
\end{align} 
where the order of the diagrams corresponds to the order of the terms in line~\eqref{eq-secondmomentrelaxation}.

Hence, the relaxation of the first and second factorial moments equals\begin{align}
\mathbb{E}[N(t)|N(0)=0]=&\mathbb{E}[N]\left(1-e^{-rt}\right)\\
\mathbb{E}[(N(t))_2|N(0)=0]=&\mathbb{E}[(N)_2]\left(1-e^{-rt}\right)^2.
\end{align}
Added together, they equal the second moment, which is depicted in Fig.~\ref{fig-moments} (solid lines).

For higher moments, the Feynman diagrams becomes quite complex, and it is advantageous to use the correspondence between cumulants and connected diagrams, similar to Sec.~\ref{sec-cumulants}. The result is that the factor $(1-e^{-rt})^n$ is attached to the $n$th steady state factorial moment to obtain the one of the relaxation from the empty system\begin{align}
\mathbb{E}[(N(t))_n|N(0)=0]=&\mathbb{E}[(N)_n]\left(1-e^{-rt}\right)^n,
\end{align} 
which is verified with simulations for the 3rd and 4th moment in Fig.~\ref{fig-moments} using Eq.~\eqref{eq-fact-moment-sterling}.

Using the relation between factorial moments and the probability distribution $P(N=\ell)$ in Eq.~\eqref{eq-steadystateprobability1}, the relaxation of the probability distribution can be calculated as well\begin{align}
P(N=\ell,t)=&\\
=\frac{\Gamma\left(\frac{\gamma}{q_2}+\ell\right)}{\ell!\Gamma\left(\frac{\gamma}{q_2}\right)}&\hspace{-0.13cm}\left(\frac{r}{r+(1-e^{-rt})q_2}\right)^{\frac{\gamma}{q_2}}\hspace{-0.15cm}\left(\frac{(1-e^{-rt})q_2}{r+(1-e^{-rt})q_2}\right)^\ell\hspace{-0.2cm}.\notag
\end{align}
It is shown for example parameters in Fig.~\ref{fig-probabilities}.

The set up described here can be generalized to initialize the system in any state or distribution of states and calculate its relaxation towards the steady state.  If the system should start at $t=0$ in state $N=\ell$, then the $n$th factorial moment at time $t$ equals $\langle\phi(t)^n\phi(0)^{\dagger \ell}e^{-\widetilde\phi(0)\phi(0)}\rangle$, where $\phi^\dagger=\widetilde\phi+1$. Initial distributions of states are then achieved by using the linearity of $\langle\bullet\rangle$.

\subsubsection{Conclusion of Relaxation}
The previous discussions on injecting particles into a steady state system, auto-correlation functions and initialization of an empty system show that the relaxation of the pumped branching process follows the function $e^{-rt}$. Other time scales of the system, such as $q_j$ or $\gamma$, do not enter. If the brain is modelled well by a pumped branching process, then a critical brain with $r=0$ would not be able to switch between different steady states because the relaxation would take an infinite amount of time. If the system is close to criticality with $\mathbb{E}[K]\approx1^-$, then the time scale of relaxation still depends on the branching time scale of the system $s$ because $r=s(1-\mathbb{E}[K])$, which is discussed based on experimental evidence in Sec.~\ref{sec-ISI-extinction-time}.

\section{Relation to experimental data}\label{sec-experiment}
Although branching processes are commonly used models for neuronal activity \cite{Beggs2003,Beggs2004,Priesemann2009,Priesemann2013,Wilting2018b}, they are not directly observed in the brain. The implanted electrode arrays detect spikes from action potentials in neurons which are interpreted as particle creations in branching models.  Thus, branching-like processes are generated by attaching time bins $\Delta t$ to spikes, where $\Delta t$ represents the expected  extinction time of a particle, Fig.~\ref{fig-schematic}.  The typical choice for the time bin is the average  inter-spike interval $\langle\text{ISI}\rangle$ of the spike time series. Whether $\langle\text{ISI}\rangle$ is the correct choice and whether the resulting process is really a branching process can be answered by reversing the procedure, i.e. by deriving the spike time series from a branching process, which is done in the following. 

In this context, events are the discrete time instances when either a particle is created or goes extinct. Creation events include both  branching and spontaneous creations but not extinctions. They will be called spike events. 

\subsection{Event distribution}\label{sec-enter-state-n}

\subsubsection{Probability to enter state $N=n$}
Given a randomly chosen event in the steady state, how likely is it that state $N=n$ is entered at this event? 

While $P(N=n)$, Eq.~\eqref{eq-steadystateprobability3}, is defined as the distribution of states $N=n$ at an \textit{arbitrary time} $t$, let $f(N=n)$ be defined as the distribution of states that the system is entering at an \textit{arbitrary event}. 

\begin{figure}
\begin{center}
\includegraphics[width=\columnwidth]{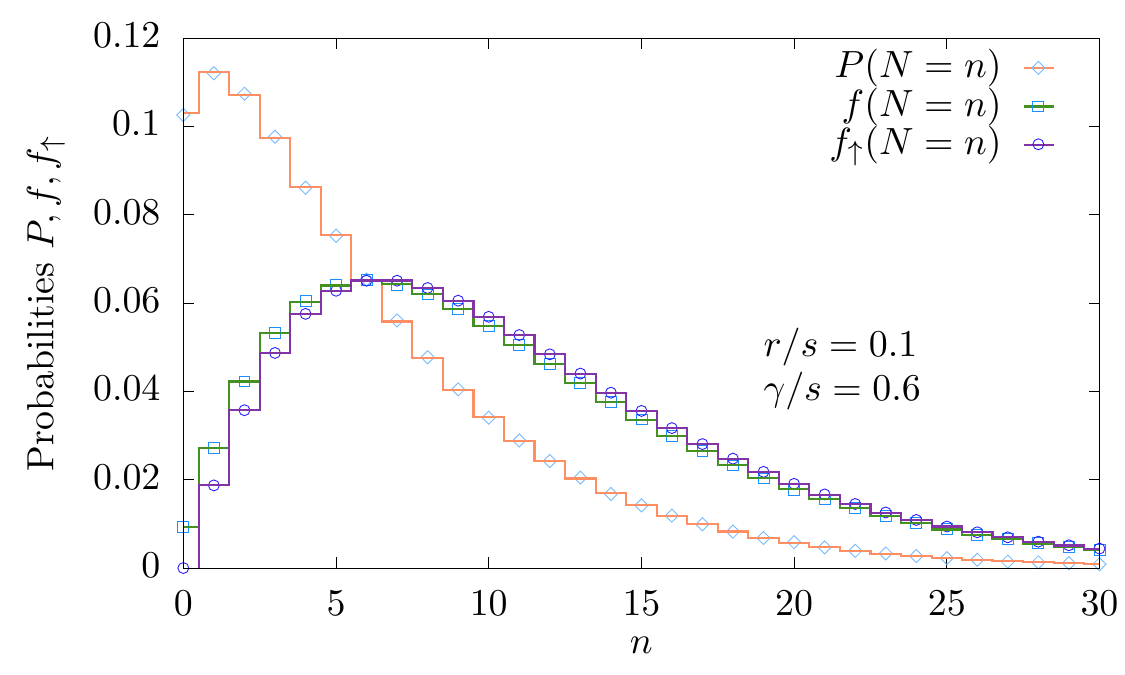}
\end{center}
\caption{Comparison of steady state probabilities. Probability $P(N=n)$ to be in state $N=n$ at any time, probability $f(N=n)$ to enter state $N=n$ at a creation or extinction event, and probability $f_\uparrow(N=n)$ to enter state $N=n$ at a creation event. The parameter values are $r/s=0.1$, $\gamma/s=0.6$. Symbols: simulation results.}\label{fig-frequency}
\end{figure}

\begin{figure}
\begin{center}
\includegraphics[width=\columnwidth]{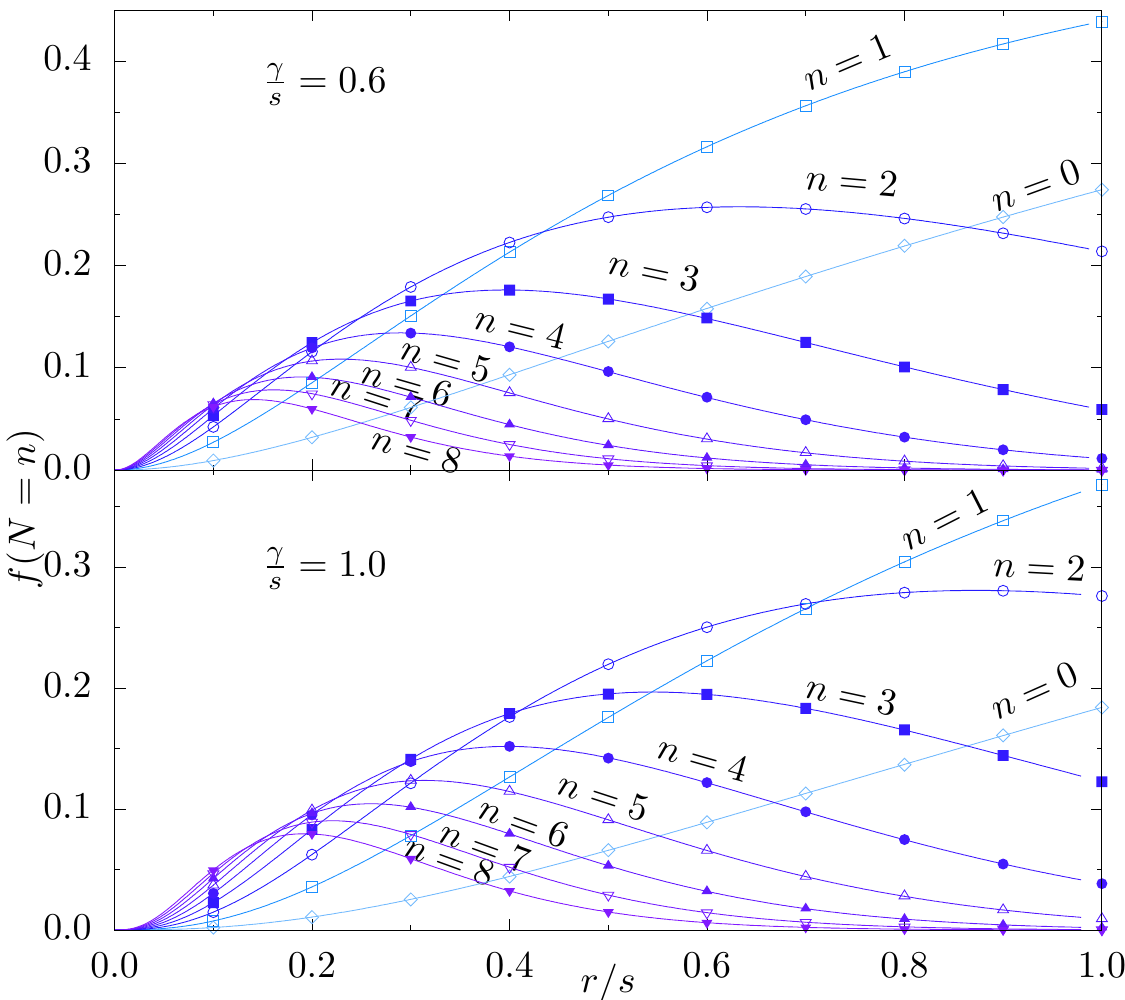}
\end{center}
\caption{Probability $f(N=n)$ that at a given event, state $N=n\in\{0,1,\dots,8\}$ is entered. Domain: $r/s\in[0,1]$. The parameter values are $\gamma/s\in\{0.6,1.0\}$. Symbols: simulation results. Lines: Analytical prediction.}\label{fig-state-entering-distribution}
\end{figure}

The difference between the two distributions is due to the fact that the system stays on average different amounts of time in different states. On average, it leaves states of large $n$ quicker compared to states with small $n$. More explicitly, if the system is in state $N=n$, it is expected to stay in there for $\frac{1}{sn+\gamma}$ amount of time, because the waiting time to the next event is exponentially distributed with rate $sn+\gamma$. This can be used to link $P(N=n)$ to $f(N=n)$ by realizing that the expected waiting time in state $N=n$ acts as a probabilistic weight:\begin{align}
\frac{P(N=n)}{P(N=m)}=\frac{f(N=n)(sm+\gamma)}{f(N=m)(sn+\gamma)}.
\end{align}
Thus, $f(N=n)$ can be written for all $n$ in terms of $P(N=n)$, $P(N=0)$ and $f(N=0)$. Then, $f(N=0)$ can be determined by imposing normalization on $f(N=n)$ and the result is
\begin{align}
f(N&=n)=\frac{1}{\gamma}\frac{r}{r+s}\left(sn+\gamma\right)P(N=n)\label{eq-state-entering-probability}\\
=&\frac{1}{\gamma}\frac{r}{r+s}\left(sn+\gamma\right)\frac{\Gamma\left(\frac{\gamma}{q_2}+n\right)}{n!\Gamma\left(\frac{\gamma}{q_2}\right)}\left(\frac{r}{r+q_2}\right)^{\frac{\gamma}{q_2}}\left(\frac{q_2}{r+q_2}\right)^{n}.
\notag
\end{align}
It is shown in Fig.~\ref{fig-frequency} and Fig.~\ref{fig-state-entering-distribution} for example parameters together with simulations as verification.

\subsubsection{Moments of the inter-event interval}
If the system is in state $N=n$, then the time to leave this state is exponentially distributed with rate $(sn+\gamma)$. However, the system does not enter all states equally often, which is captured by $f(N=n)$, Eq.~\eqref{eq-state-entering-probability}. Therefore, the time $T_e$ between events is in expectation\begin{align}
\mathbb{E}[T_e]=&\sum\limits_{n=0}^\infty f(N=n) \int\limits_{0}^\infty t(sn+\gamma)e^{-(sn+\gamma)t}\\
=&\frac{1}{\gamma}\frac{r}{r+s}.
\end{align}
Unfortunately, higher moments do not have such simple expressions, and the $m$th moment of the event time distribution is
\begin{align}
\mathbb{E}[T_e^m]=&\frac{m!}{\gamma^m}\frac{r}{r+s}\left(\frac{r}{r+q_2}\right)^{\frac{\gamma}{q_2}}F,\\
\text{with}\hspace{1.5cm}&\notag\\
F={_mF_{m-1}}&\Bigl(\frac{\gamma}{q_2},\underbrace{\frac{\gamma}{s},\dots,\frac{\gamma}{s}}_{m\text{ times}};\underbrace{1+\frac{\gamma}{s},\dots,1+\frac{\gamma}{s}}_{m\text{ times}};\frac{q_2}{r+q_2}\Bigr),\notag
\end{align}
where ${_mF_{m-1}}$ is a hypergeometric function. It is verified using simulations in Appendix~\ref{sec-interval-verification} in Fig.~\ref{fig-event-time-moments} for the 1st, 2nd and 3rd moment of $T_e$.

\subsubsection{Event interval distribution $f_e(t)$}
If a time is picked at random, how is the waiting time to the next event distributed? At the observation time, the system is in state $N=n$ with probability $P(N=n)$. Within this state, the system is memoryless, i.e. it does not know how long it has been in there. Hence, the remaining event time before the next event is exponentially distributed with probability density $(sn+\gamma)e^{-(sn+\gamma)t}=f_e(t|n)$. Thus, $f_e$ can be recovered by summing $f_e(t|n)P(N=n)$ over $n$, which gives\begin{align}
f_e(t)=\gamma e^{-\gamma t}\frac{(r+q_2)(1-e^{-st})}{r+(1-e^{-st})q_2}\left(\frac{r}{r+(1-e^{-st})q_2}\right)^\frac{\gamma}{q_2}.
\end{align}
The details of this calculation can be found in Appendix~\ref{sec-event-time-distribution}.
However, inter-event times are not independent of each other and therefore $f_e(t)$ should not be used for subsequent events unless the amount of data is large enough such that the correlations between subsequent event times are averaged out.

\subsection{Spike distribution}\label{sec-spike-time-distribution}
The recordings of neuronal avalanches consists of time series of spikes. If spikes occur every time a neuron becomes active, then -- translated into this model --  spikes occur every time a new particle is created, either through branching or spontaneous creation.  Therefore, the first step is the derivation of the distribution of spike events within all events. As before, binary branching is assumed, i.e. the offspring distribution is given by $p_0$ and $p_2$ and the extinction time of a single particle is exponentially distributed with rate $sp_0$. 

\subsubsection{Probability to enter state $N=n$ via creation}
The distribution $f(N=n)$, Eq.~\eqref{eq-state-entering-probability}, describes how likely it is to enter state $N=n$ at an arbitrary event, either through creation or extinction. We want to filter out the extinction events and therefore define the probability $f_\uparrow(N=n)$ to enter state $N=n$ through a creation event. It is found by considering that state $N=n-1$ is left  via a creation with probability $\frac{sp_2(n-1)+\gamma}{s(n-1)+\gamma}$, while it is left through an extinction with probability $\frac{sp_0(n-1)}{s(n-1)+\gamma}$. Therefore, $f_\uparrow(N=n)$ is proportional to $f(N=n-1)$ multiplied by the chance that state $n-1$ is followed by a creation event. The result has to be normalized and is found to be\begin{align}
f_\uparrow(N=n)=2f(N=n-1)\frac{sp_2(n-1)+\gamma}{s(n-1)+\gamma},
\end{align}
i.e. the normalization constant equals $2$. $f_\uparrow$ is shown for example parameters in Fig.~\ref{fig-frequency} alongside simulations. 

\subsubsection{Moments of the inter-spike interval}

The time between spikes $T_s$  is a random variable, often referred to as inter-spike interval (or ISI). Its expected value $\mathbb{E}[T_s]$ is commonly denoted by $\langle\text{ISI}\rangle$, which will be used interchangeably here. $T_s$ is \textit{not} exponentially distributed, because the system can traverse several states via a sequence of extinctions between spikes. For example, if the system is in state $N=m$, then it can go to state $m-1$, then $m-2$, and so on, before a particle is created and a spike occurs. The durations between the extinctions are each exponentially distributed with slightly different rates. The overall time is the sum of exponentially distributed random variables and is therefore \textit{hypo}exponentially distributed. However, within this broad family of distributions, $T_s$ has a specific structure because the summed exponential random variables have rates $sm+\gamma$, $s(m-1)+\gamma$, $\dots$. This allows for a slightly more elegant way of writing moments of $T_s$ than in the case of arbitrary hypoexponential distributions.  

Given that the system is in state $N=m>0$, the $n$th moment of the waiting time $T_s$ to the next spike  equals\begin{align}
\mathbb{E}[T_s^n&|N=m]=\label{eq-spike-waiting-time-from-fixed-state}\\
=&\sum\limits_{\ell=0}^m (\ell sp_2+\gamma)\left(-\frac{\plaind}{\plaind\gamma}\right)^n\left(\frac{1}{\ell s+\gamma}\prod\limits_{k=\ell+1}^m \frac{ksp_0}{ks+\gamma}\right).\notag
\end{align}
Its derivation is explained in Appendix~\ref{appendix-spike-waiting-time-from-fixed-state}.

Hence, the moments of the inter-spike interval $T_s$ can be calculated by summing over initial states $N=m$ weighted by the likelihood $f_\uparrow(N=m)$ to enter this state through a creation:\begin{align}\label{eq-spike-waiting-time}
\mathbb{E}[T_s^n]=\sum\limits_{m=1}^\infty f(m-1\uparrow m)\mathbb{E}[T_s^n|N=m],
\end{align}
which does not appear to have a simpler expression. However, it can be evaluated numerically  to high precision in a large parameter region. A program for calculating moments of inter-spike intervals in a large parameter region is available at \cite{Pausch2020b}. The theory is verified with simulations in Appendix~\ref{sec-interval-verification} in Fig.~\ref{fig-spike-time-moments}.

\begin{figure}
\begin{center}
\includegraphics[width=\columnwidth]{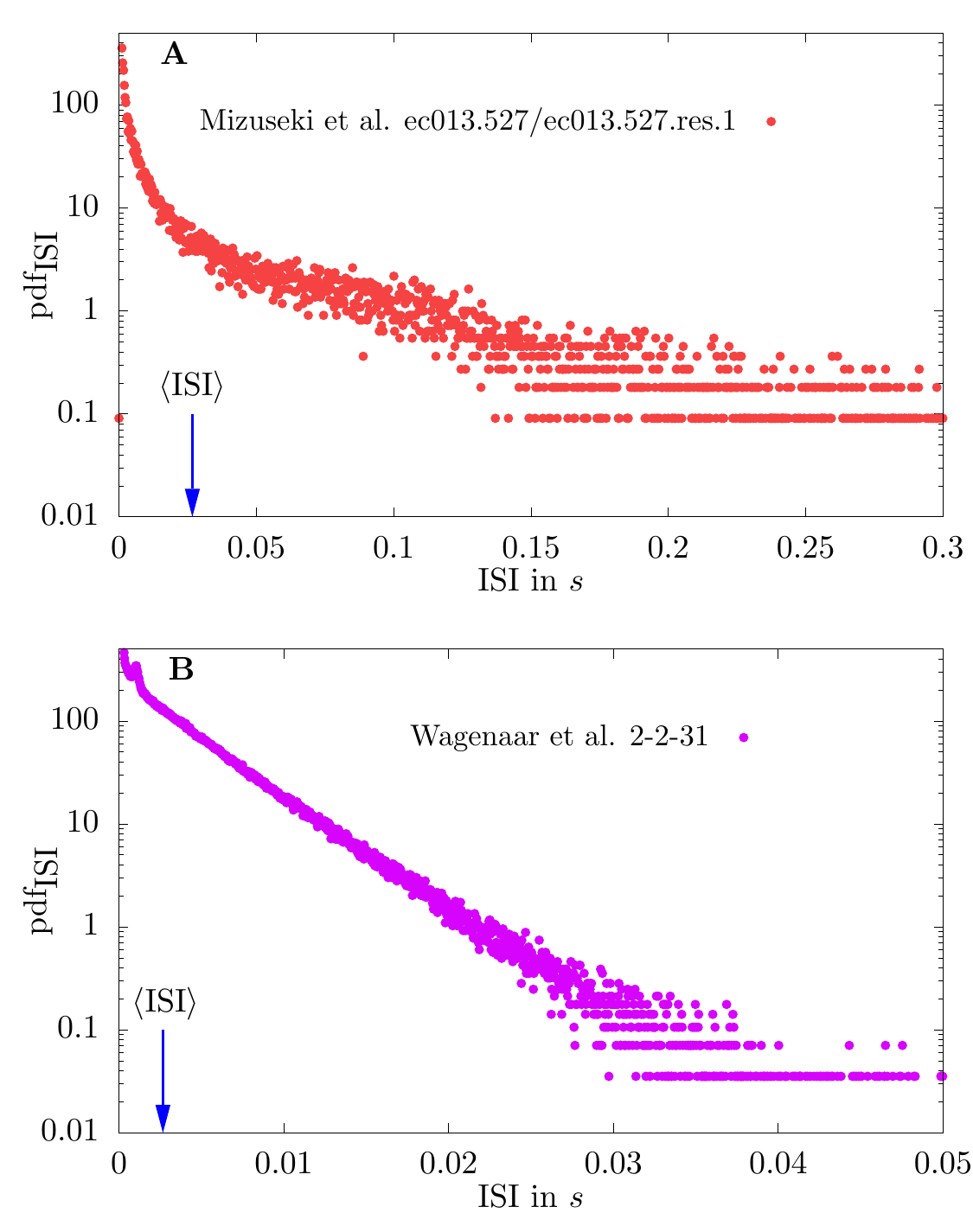}
\caption{Estimated probability density function pdf$_\text{ISI}$ of the inter-spike interval distributions. \textbf{A}: data set ec013.527/ec013.527.res.1 from  \cite{Buzsaki2009}. \textbf{B}  data set 2-2-31 from \cite{Wagenaar2006}. Both show a continuous distribution of the ISI. The approximate straight line on the  logarithmic scale in Panel \textbf{B} suggests that an exponential distribution might be a good approximation to the distribution.}
\label{fig-EXP-inter-spike-dsitribution}
\end{center}
\end{figure}

Now that the moments of the inter-spike interval are derived, they can be compared to experimental data. In the following, two sources for experimental data are used. \begin{enumerate}
\item The first data sets are available from \url{crcns.org} (data hc-2, 124 data sets) and were published by Mizuseki et al.~(2009) \cite{Buzsaki2009}. The data was collected \textit{in vivo} from the CA1 layer of the right dorsal hippocampus of three rats, with identifiers ec013 to ec016 (ec014 and ec015 denote the same rat), using probes consisting of 4 or 8 shanks, with 8 recording sites per shank. A more detailed description can be found in Appendix~\ref{app-data-description}, which is a reproduction of its description on crcns.org. Full details of the experimental set up can be found in \cite{Buzsaki2009}. In Figs.~\ref{fig-EXP-inter-spike-dsitribution}, \ref{fig-coefficient-of-variation}, \ref{fig_SpikeMomentsMapExpData}, \ref{fig-ISI-vs-extinction-time}, \ref{fig-avalanche-size-length-time-bin} and~\ref{fig_Approximation_Error}, the data is shown as red symbols. In Fig.~\ref{fig-avalanche-size-distribution}, the data is shown as both red and purple symbols.
\item The second data sets are available from \url{neurodatasharing.bme.gatech.edu/development-data/} and were published by Wagenaar et al.~(2006) \cite{Wagenaar2006}. The data was collected \textit{in vitro} from cells plated on multi-electrode arrays with 59 electrodes. Data was recorded for up to 39 days after plating ($3\le$ days in vitro (div) $\le 39$). The dissociated cells were obtained from rat embryos. A short description of the data collection procedure can be found in the Appendix~\ref{app-data-description}. All 527 data sets labelled as `dense' are used in the following. Details of the experimental setup can be found in \cite{Wagenaar2006}. In Figs.~\ref{fig-EXP-inter-spike-dsitribution}, \ref{fig-coefficient-of-variation}, \ref{fig_SpikeMomentsMapExpData}, \ref{fig_ComparisonContinuousDiscrete}, \ref{fig-ISI-vs-extinction-time}, and~\ref{fig_Approximation_Error}, the data is shown as purple symbols. 
\end{enumerate}

One major premise of the proposed model was that inter-spike times are continuously distributed. This can be verified by considering examples from the above data sets. Two examples are shown in Fig.~\ref{fig-EXP-inter-spike-dsitribution}. Both examples clearly suggest a continuous distribution of the ISI and Panel \textbf{B} in Fig.~\ref{fig-EXP-inter-spike-dsitribution} even strongly suggests an underlying exponential distribution.

\subsection{Coefficient of variation}\label{sec-CV}
The coefficient of variation $c_V$ is commonly used in the analysis of spike time series in neuroscience.\cite{Pfeiffer1965,Nakahama1968,Lamarre1971,Bassant1976,Rinzel1983,Softky1993,Fontenele2019} It is defined as\begin{align}
c_V=&\frac{\sigma_{T_s}}{\mathbb{E}[T_s]},
\end{align}
where $\sigma_{T_s}$ is the standard deviation of the inter-spike interval $T_s$.

\begin{figure}
\begin{center}
\includegraphics[width=\columnwidth]{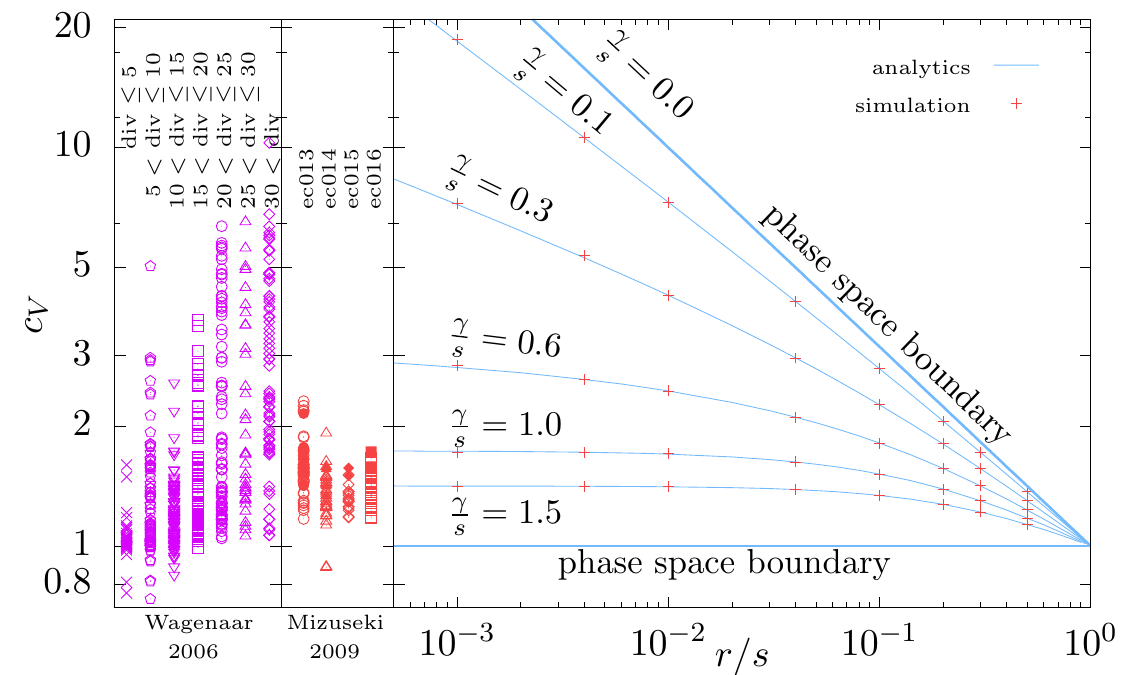}
\caption{Coefficient of variation $c_V$. Right: Analytical predictions (solid lines) and simulation results (plus symbols) over $\frac{r}{s}$ for several values of $\frac{\gamma}{s}$. Left: experimental data from \cite{Buzsaki2009} (red) and \cite{Wagenaar2006} (purple). For \cite{Buzsaki2009}, solid symbols are data from combined time series of all shanks of the probe, hollow symbols are data from individual shanks of the probe. }
\label{fig-coefficient-of-variation}
\end{center}
\end{figure}

The coefficient of variation $c_V$ is dimensionless and only depends on $\frac{r}{s}$ and  $\frac{\gamma}{s}$. It shows a phase space boundary across which the pumped branching process cannot explain the spike statistics. This boundary is found when $\gamma\rightarrow0$, i.e. when the clustering coefficient $\frac{\gamma}{q_2}\rightarrow0$ indicates that the process exhibits concentrated bursts followed by long periods without spikes. The boundary can be calculated analytically to be equal to $c_V(\frac{r}{s},\frac{\gamma}{s}=0)=\sqrt{\frac{s}{r}}$. The derivation of this result is explained in Appendix~\ref{sec-phase-space-boundary}. A second phase space boundary is found in the limit of $\frac{\gamma}{s}\rightarrow\infty$, for which the process becomes a Poisson process. 

However, $c_V$ is a scalar function depending on two variables and is therefore not invertible, as shown in Fig.~\ref{fig-coefficient-of-variation} together with experimental data.\cite{Buzsaki2009,Wagenaar2006} The figure shows that the vast majority of data sets cover a range of $1<c_V<10$, with few outliers outside this range. However, this can be reproduced by a wide range of spike time series induced by pumped branching processes. For the \textit{in vivo} data (Mizuseki et al.~(2009)), the small gap between the $c_V$ from experiments and $c_V=1$ allows only to exclude $\frac{r}{s}\gtrapprox0.9$ for the entire data set. However, the largest $c_V$ values in the data set can only be explained by degrees of criticality $\frac{r}{s}<0.2$. The \textit{in vitro} data (Wagenaar et al.~(2006)) shows an increase in $c_V$ over the 5 weeks after plating of the cells in line with the reported increase in burstiness reported in \cite{Wagenaar2006}.  

In order to better understand the link between spike data and pumped branching processes, an invertible mapping needs to be found.

\subsection{The moment-ratio map}\label{sec-moment-ratio-map}
In order to achieve a quantitative comparison to experimental data from neuronal circuits, moment ratios are considered in the following. Defining\begin{align}\label{eq-moment-ratios}
X=&\frac{\mathbb{E}[T_s^3]}{\mathbb{E}[T_s]^3}-6\\
Y=&\frac{\mathbb{E}[T_s^4]}{\mathbb{E}[T_s^2]^2}-6
\end{align}
allows eliminating the explicit dependence on parameter $s$, which defines a time scale for the stochastic process. It will only appear in dimensionless combinations of $\frac{r}{s}$ (degree of criticality) or $\frac{\gamma}{s}$ (relative spontaneous creation). The definition also implies that the variables $X$ and $Y$ are dimensionless. The shift by $-6$ is introduced to position the Poisson process without branching at the origin. 

\begin{figure}
\begin{center}
\includegraphics[width=\columnwidth]{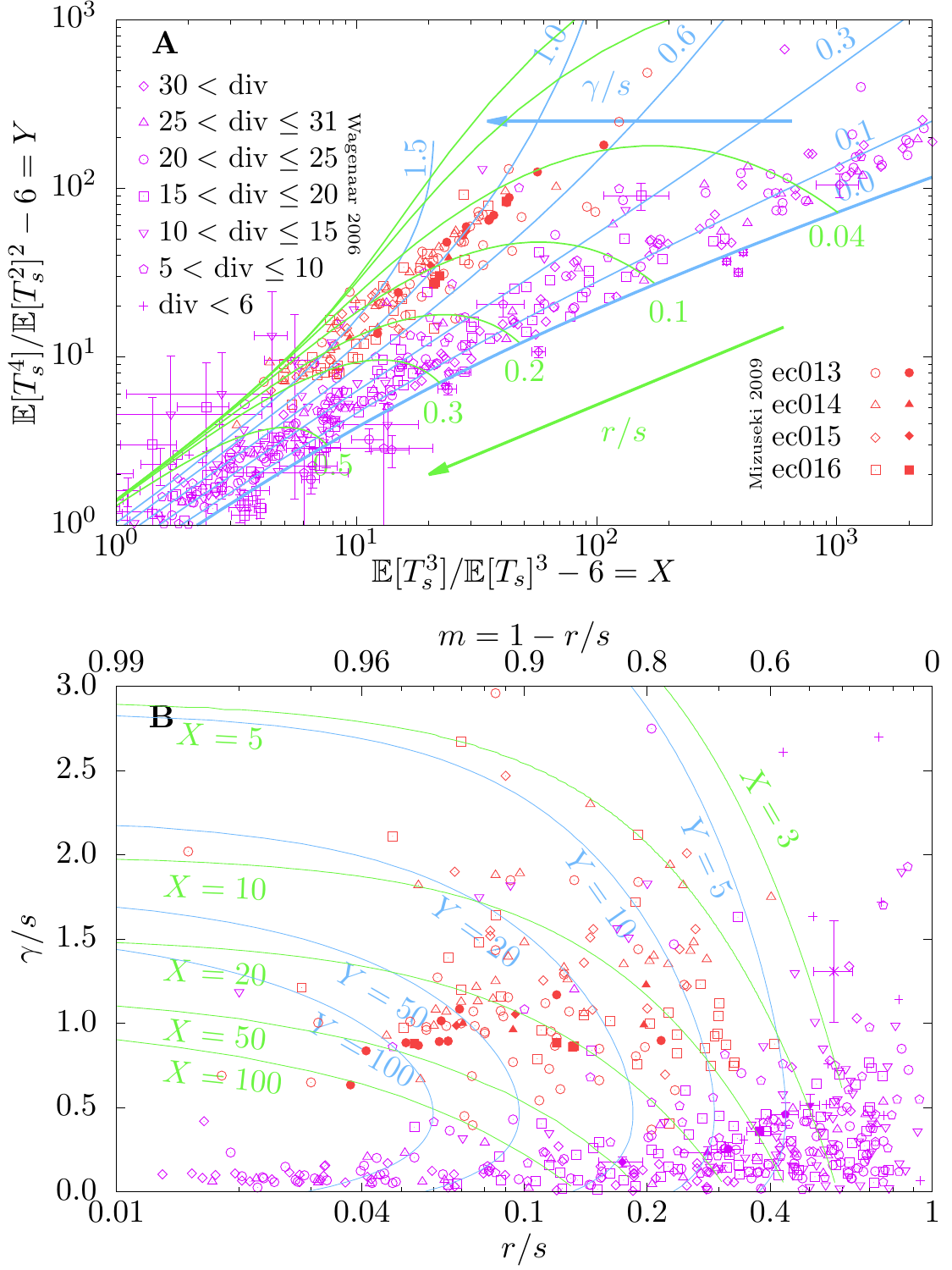}
\caption{Moment-ratio map. \textbf{A}: Blue lines:  $\frac{\gamma}{s}$-level sets, green lines:  $\frac{r}{s}$-level sets. The area below the thick blue line $\frac{\gamma}{s}=0$ is not in the phase space of the spike process induced by pumped branching processes.  \textbf{B}: Experimental data mapped onto $\frac{r}{s}$-$\frac{\gamma}{s}$ space (and $m$-$\frac{\gamma}{s}$ space), blue lines: $Y$ level sets, green lines: $X$ level sets. Both \textbf{A} and \textbf{B}: Red symbols: \textit{in vivo} experimental data from \cite{Buzsaki2009} that used four rats with identifiers ec013 to ec016, using a total of 124 data sets. Solid red symbols are data from combined time series of all shanks of the probe, hollow symbols are data from individual shanks. Purple symbols: experimental data from \cite{Wagenaar2006}, using all 527 data sets labelled as `dense'. Different symbol shapes show different days \textit{in vitro}. Solid purple symbols in \textbf{B}: average over time periods according to shape of symbol.}
\label{fig_SpikeMomentsMapExpData}
\end{center}
\end{figure}

The moment ratios $X$ and $Y$ can be used to create the map in Fig.~\ref{fig_SpikeMomentsMapExpData}\textbf{A}, which shows levels sets of $\frac{r}{s}$ and $\frac{\gamma}{s}$ in an $X$-$Y$ graph. The map itself is verified with Monte Carlo simulations in Appendix~\ref{sec-verification-ratio-map}, Fig.~\ref{fig_SpikeMomentsMapVerification}. Fig.~\ref{fig_SpikeMomentsMapExpData}\textbf{A} also shows experimental data \cite{Buzsaki2009,Wagenaar2006}. In order to avoid a too crowded plot, error bars are only shown for data outside of the computational range (left side) or outside the phase space boundary (bottom) as well as a few additional examples. The plot also shows a phase space boundary for $\frac{\gamma}{s}\rightarrow0$, which is derived in Appendix~\ref{sec-phase-space-boundary}. A phase space boundary in the direction of small $X$ and large $Y$ has not been found. Furthermore, evaluating the functions $X$ and $Y$ in that region of parameter space is computationally difficult and 128bit double precision becomes insufficient. 

The level sets in the moment-ratio map indicate that there is a 1-to-1 correspondence (or bijection) between $(X,Y)$ and $(\frac{r}{s},\frac{\gamma}{s})$ in the parameter region of interest. As the analytic inversion of the map is very difficult, it is done using numerical methods of which the result is shown in Fig.~\ref{fig_SpikeMomentsMapExpData}\textbf{B}. The procedure is as follows\begin{enumerate}
\item Use the open-source code in \cite{Pausch2020b} to calculate moments of inter-spike intervals for trial values of $\frac{r}{s}$ and $\frac{\gamma}{s}$,
\item Determine the inter-spike interval moments from the data,
\item Combine step 1 and 2 in a gradient descent method to find the unique $\frac{r}{s}$ and $\frac{\gamma}{s}$ corresponding to the data.
\end{enumerate}
The algorithmic error margin for $\frac{r}{s}$ values was allowed to be at most 0.0001, while for $\frac{\gamma}{s}$ it was at most 0.001. Fig.~\ref{fig_SpikeMomentsMapExpData}\textbf{B} shows level sets of $X$ and $Y$ as well as the experimental data. It gives a clearer picture of the spread of parameter values to which the experimental data corresponds. 

Most \textit{in vivo} data lies in a range of $\frac{r}{s}\in(0.04,0.4)$ ($m\in(0.6,0.96)$) and $\frac{\gamma}{s}\in(0.5,2.5)$ with few outliers. The different data sets from the four different rats do not appear to cluster around different parameter values. 

Most \textit{in vitro} data lies in a range of $\frac{r}{s}\in(0.02,0.9)$ ($m\in(0.1,0.98)$) and $\frac{\gamma}{s}\in(0,0.5)$ with few outliers. Although the degree of criticality of the \textit{in vitro} data is widely spread, there is a trend that the degree of criticality $\frac{r}{s}$ increases the longer the cells have been plated. In order to quantify this trend, the averages of each time period (see legend in Fig.~\ref{fig_SpikeMomentsMapExpData}\textbf{A}) are shown as solid symbols with error bars. The averages almost perfectly line up to show an increase of the degree of criticality of $\frac{r}{s}\approx0.6$ ($m\approx0.4$) for div $\le5$ to $\frac{r}{s}\approx0.18$ ($m\approx0.82$) for div $>30$, which confirms the findings in \cite{Levina2017}. This change can be interpreted as the increasing influence of branching in the developing neural network.

Comparing \textit{in vivo} with \textit{in vitro} data supports the findings in \cite{Zierenberg2018} that different external input (i.e.~pumping / immigration) explains the different observed spiking dynamics.

The identified parameters can be used to estimate the range of the number of active neurons during steady state activity. For example for the \textit{in vivo} data, the average particle number in the system is $\frac{\gamma}{r}$, Eq.~\eqref{eq-firstmoment}, hence the average  number of active neurons in the capture area of electrode arrays ranges from 60 active neurons to 1 active neuron at any given time, depending on where the data set lies in the $\frac{r}{s}$-$\frac{\gamma}{s}$ space. 

In addition, the found parameters allow estimating the avalanche size and the average number of causal avalanches \cite{Munoz2017} within one avalanche, see Eq.~\eqref{eq-expected-length} and~\eqref{eq-expected-size}. As an example, let's look at the combined data set of ec016.448 (\textit{in vivo}), which contains the superimposed data of eight shanks. Its determined parameters are a degree of criticality of $\frac{r}{s}=0.13125$ ($m=0.86875$), and a relative spontaneous creation of $\frac{\gamma}{s}=0.86$, which imply an average size of $\approx78$ spikes per avalanche and an average of $\approx$ 17 causal avalanches within one avalanche, each containing an average of 4 to 5 spikes. Hence for this example, the common separation-of-time-scales (STS) assumption is not valid. As a different example, we consider the \textit{in vitro} data set 7-2 for div 33 which has $\frac{r}{s}=0.01953$ ($m=0.98047$) and $\frac{\gamma}{s}=0.11$. This implies that avalanches contain on average $\approx54$ spikes and are made up of $\approx1$ causal avalanche. Hence, for this example, the STS assumption makes sense.

Once the parameter values $\frac{r}{s}$ and $\frac{\gamma}{s}$ have been identified, it can be checked how accurately the theoretical  spike interval distribution models the observed one by comparing other observables $\mathcal{O}$. This is done in Appendix~\ref{sec-approx-error}, showing approximation errors of only a few percent for a variety of observables. 

\begin{figure}
\begin{center}
\includegraphics[width=\columnwidth]{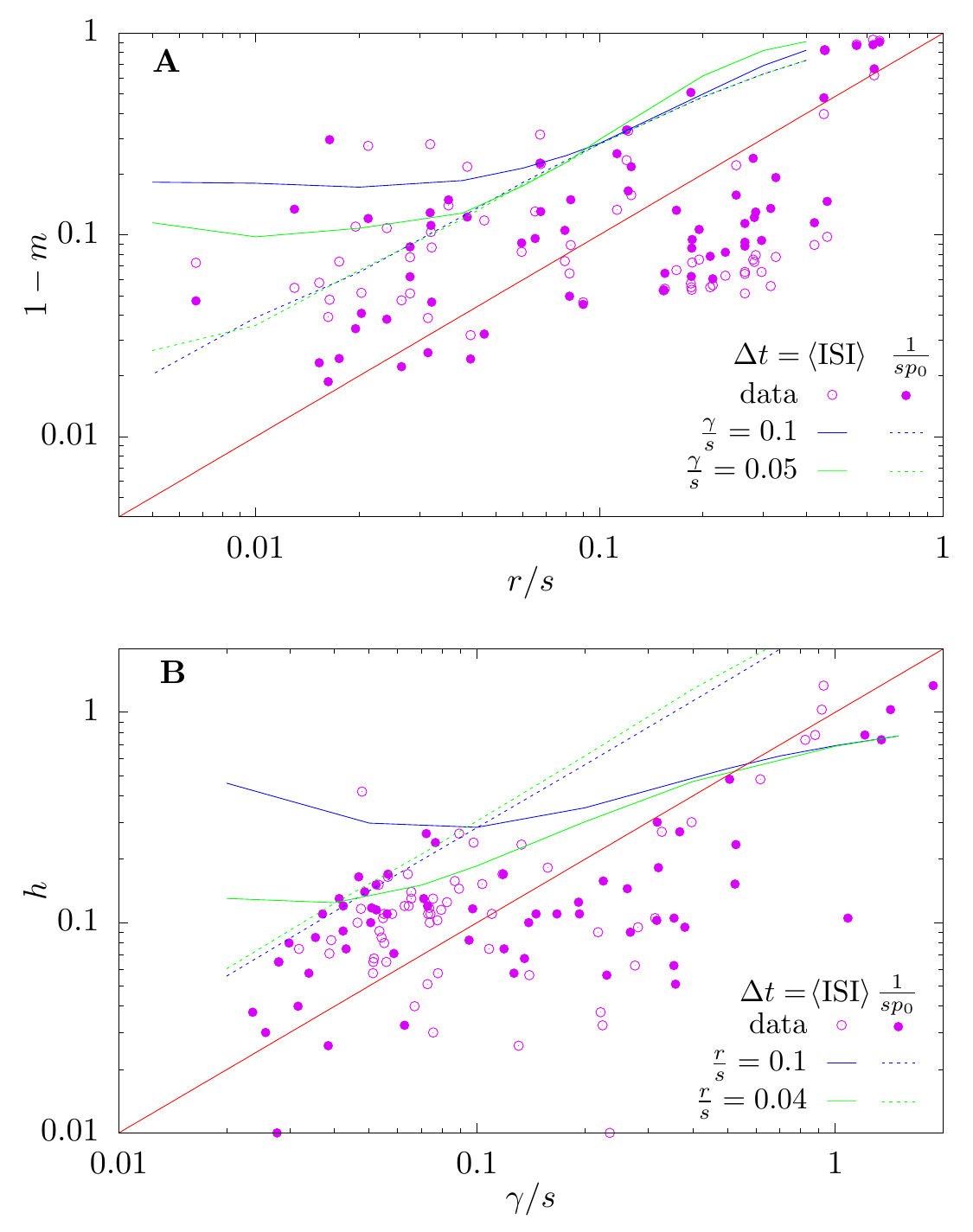}
\caption{Parameter comparison between continuous and discrete branching process with immigration / pumping. All the data sets from \cite{Wagenaar2006} for div $>30$ are shown as symbols. For the discrete process a choice of time-step $\Delta t$ had to be made: hollow circles represent $\Delta t=\langle\text{ISI}\rangle$ (ISI is the inter-spike interval), solid circles represent $\Delta t=\frac{1}{sp_0}$ where $s$ and $p_0$ were inferred from the continuous process. See Sec.~\ref{sec-ISI-extinction-time} for a discussion of this choice. The parameters $m$ and $h$ were inferred by linear regression equal to the one-step linear regression in \cite{Wilting2018b}. The blue and green lines are the results when the simulated continuous-time process is discretised using time bins and then the regression analysis is applied to the resulting time series.  \textbf{A}:  comparison of the degree of criticality $\frac{r}{s}$ of continuous process with the branching parameter of the discrete process $m$. The straight line indicates the case of a hypothetical exact correspondence. \textbf{B}: comparison of the relative spontaneous creation $\frac{\gamma}{s}$ with the immigration probability $h$ of the discrete process. The straight red line indicates the case of a hypothetical exact correspondence. }
\label{fig_ComparisonContinuousDiscrete}
\end{center}
\end{figure}

The degree of criticality $\frac{r}{s}$ and the relative spontaneous creation $\frac{\gamma}{s}$ were calculated based on the assumption that the observed spike process is based on a \textit{continuous-time} pumped branching process. So far in the literature, its discrete-time version, the branching process with immigration, has been used more often \cite{Wilting2018b}. The parameters of the continuous process have a natural correspondence to the parameters of the discrete one: $\frac{r}{s}$ corresponds to $1-m$ and $\frac{\gamma}{s}$ corresponds to $h$. Both $m$ and $h$ can be inferred by (multi-step) linear regression \cite{Wilting2018b} and a comparison of the resulting parameters is shown in Fig.~\ref{fig_ComparisonContinuousDiscrete}. This figure also shows what happens when the linear regression analysis from  \cite{Wilting2018b} is applied to a pumped branching process (blue and green lines): the discretisation of time by time bins introduces a bias visible as the vertical distance between the blue/green lines and the red line.

\begin{figure}
\begin{center}
\includegraphics[width=\columnwidth]{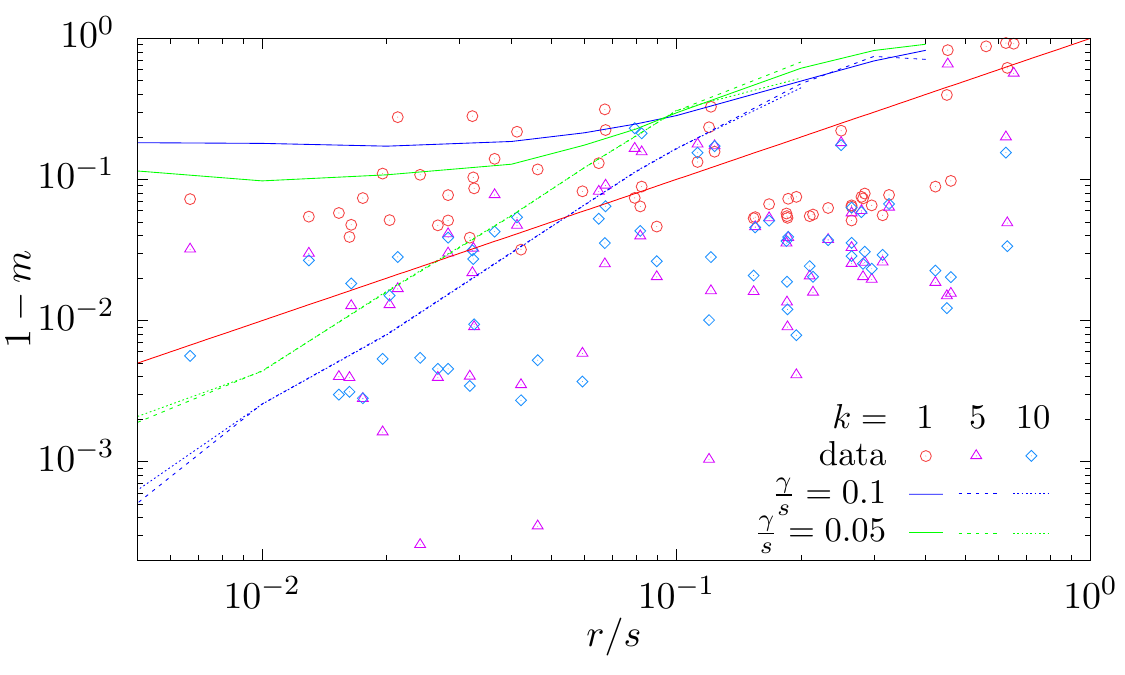}
\caption{Parameter comparison between continuous and discrete branching process with immigration / pumping. All the data sets from \cite{Wagenaar2006} for div $>30$ are shown as symbols. The blue and green lines are the results when the simulated continuous-time process is discretised using time bins and then the (multi-step) regression analysis is applied to the resulting time series. For the discrete process the choice of time-step is $\Delta t=\langle\text{ISI}\rangle$. The parameter $m$ was inferred by a multi-step linear regression proposed in \cite{Wilting2018b}. $k$ is the number of time steps considered for the regression. The straight red line indicates the case of a hypothetical exact correspondence. }
\label{fig_ComparisonContinuousDiscreteMultiStep}
\end{center}
\end{figure}

The parameter $m$ of the discrete process can be determined more accurately by a multi-step regression \cite{Wilting2018b}. Here, instead of considering only one time step to determine $m$, $k$ time steps are considered together. This method is thus analysing the auto-correlation of the process. In Sec.~\ref{sec-twotimecorrelation} it was pointed out that if this analysis is applied to a pumped branching process, its estimator for $m$ is biased. Fig.~\ref{fig_ComparisonContinuousDiscreteMultiStep}  shows a comparison of the degree of criticality for several numbers of time steps for the regression. The figure shows that the multi-step regression ($k>1$) estimates higher degrees of criticality than were determined using the moment-ratio map. The figure also shows what happens when a pumped branching process is discretised by time bins and analysed using the multi-step linear regression \cite{Wilting2018b}: a bias is introduced which is visible as the vertical distance between blue/green lines and the red line.

\subsection{Time scales and avalanche size distributions}\label{sec-ISI-extinction-time}
Since the dimensionless quantities $\frac{r}{s}$ and $\frac{\gamma}{s}$ are determined using the moment-ratio map, Fig.~\ref{fig_SpikeMomentsMapExpData}, the time scale $s$ can be determined by comparing the experimental inter-spike intervals with the theoretically expected one. Once $\frac{r}{s}$ and $\frac{\gamma}{s}$ are fixed, dimensional analysis implies that $\mathbb{E}[T_s]$ scales as $\frac{1}{s}$. Hence, they are 1-to-1 and $s$ can be determined uniquely for each data set, which completes the mapping between spike time series and pumped branching processes. 

\begin{figure}
\begin{center}
\includegraphics[width=\columnwidth]{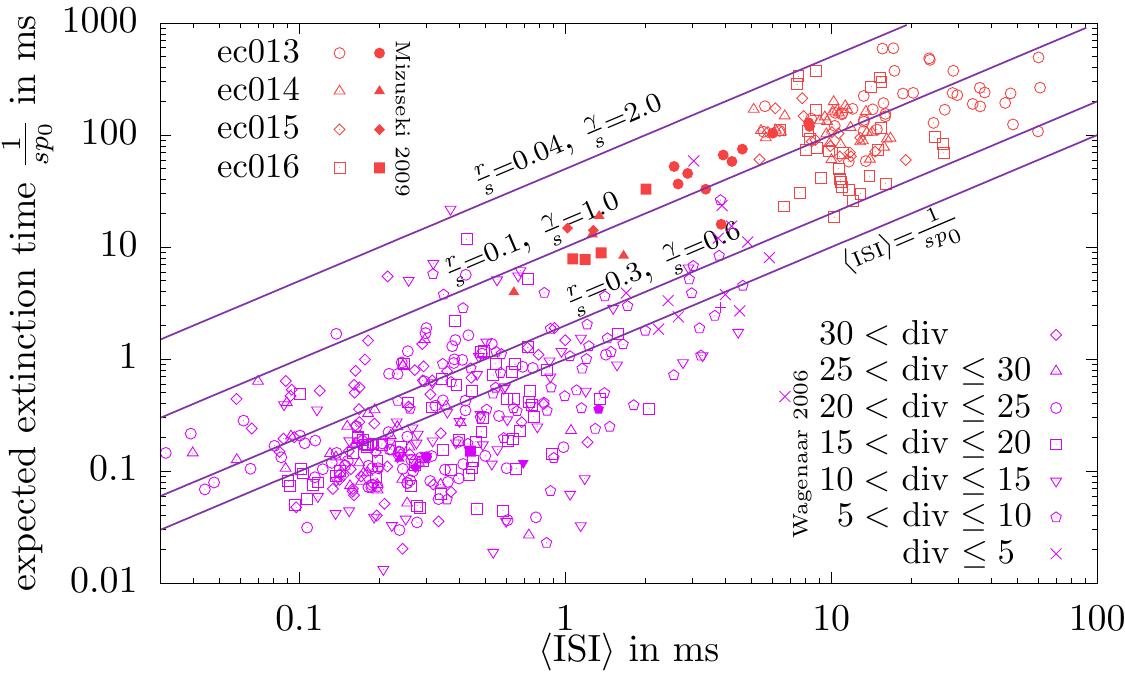}
\caption{Relation between average inter-spike interval $\langle\text{ISI}\rangle$ and expected extinction time $\frac{1}{sp_0}$ for experimental \textit{in vivo} data \cite{Buzsaki2009} (red symbols) and \textit{in vitro} data \cite{Wagenaar2006} (purple symbols). Lines: Analytical results for fixed values of degree of criticality $\frac{r}{s}$ and relative spontaneous creation $\frac{\gamma}{s}$, but varying time scale $s$. Solid red symbols show \textit{in vivo} data from combined time series of all shanks of the probe, hollow red symbols show \textit{in vivo} data from individual shanks of the probe. Solid purple symbols show averages of \textit{in vitro} data over time periods listed in legend.}
\label{fig-ISI-vs-extinction-time}
\end{center}
\end{figure}

It is now possible to answer the question whether the $\langle\text{ISI}\rangle$ is equal to the expected extinction time $\frac{1}{sp_0}$ of an activated neuron. This question is important because $\langle\text{ISI}\rangle$ is commonly used as if it were the expected extinction time to create branching-like processes.\cite{Beggs2003,Beggs2004,Priesemann2009,Priesemann2013,Lombardi2014} The answer is that in general $\langle\text{ISI}\rangle$ and expected extinction time are not equal. This discrepancy is shown for the experimental data in Fig.~\ref{fig-ISI-vs-extinction-time}. It illustrates that the bin size $\Delta t$ for recreating the underlying branching-like process should be approximately ten times as high as the $\langle\text{ISI}\rangle$ for most of the  \textit{in vivo} data. For the \textit{in vitro} data, $\langle\text{ISI}\rangle\approx\frac{1}{sp_0}$, although for the majority of data sets $\langle\text{ISI}\rangle$ is slightly larger than the expected extinction time $\frac{1}{sp_0}$, which can be explained by long inter-avalanche waiting times which increase the average ISI.  

In Sec.~\ref{sec-empty-system}, it is found that relaxation behavior in a pumped branching process system follows the function $e^{-rt}$. Although exact relaxation time scales depend on the initial and final states, an approximate scale for relaxation behavior  can be determined from the data. For most of the \textit{in vivo} data sets, the determined parameters indicate that $e^{-rt}$ would fall from 1 at $t=0$ to $0.01$ within a few seconds, not hundreds of  milliseconds and not tens of seconds. This insight opens up several questions for experimental investigation: 1) Does a switch in activity, for example between deep sleep, REM sleep, or wakefulness\cite{Priesemann2013}, rely  on changing the spontaneous creation rate $\frac{\gamma}{s}$ (i.e. the external input), or does it rely on changing the degree of criticality $\frac{r}{s}$ within the neural circuit? 2) Does the observation of the transition between steady states allow extracting these parameters of the system?     

\begin{figure}
\begin{center}
\includegraphics[width=\columnwidth]{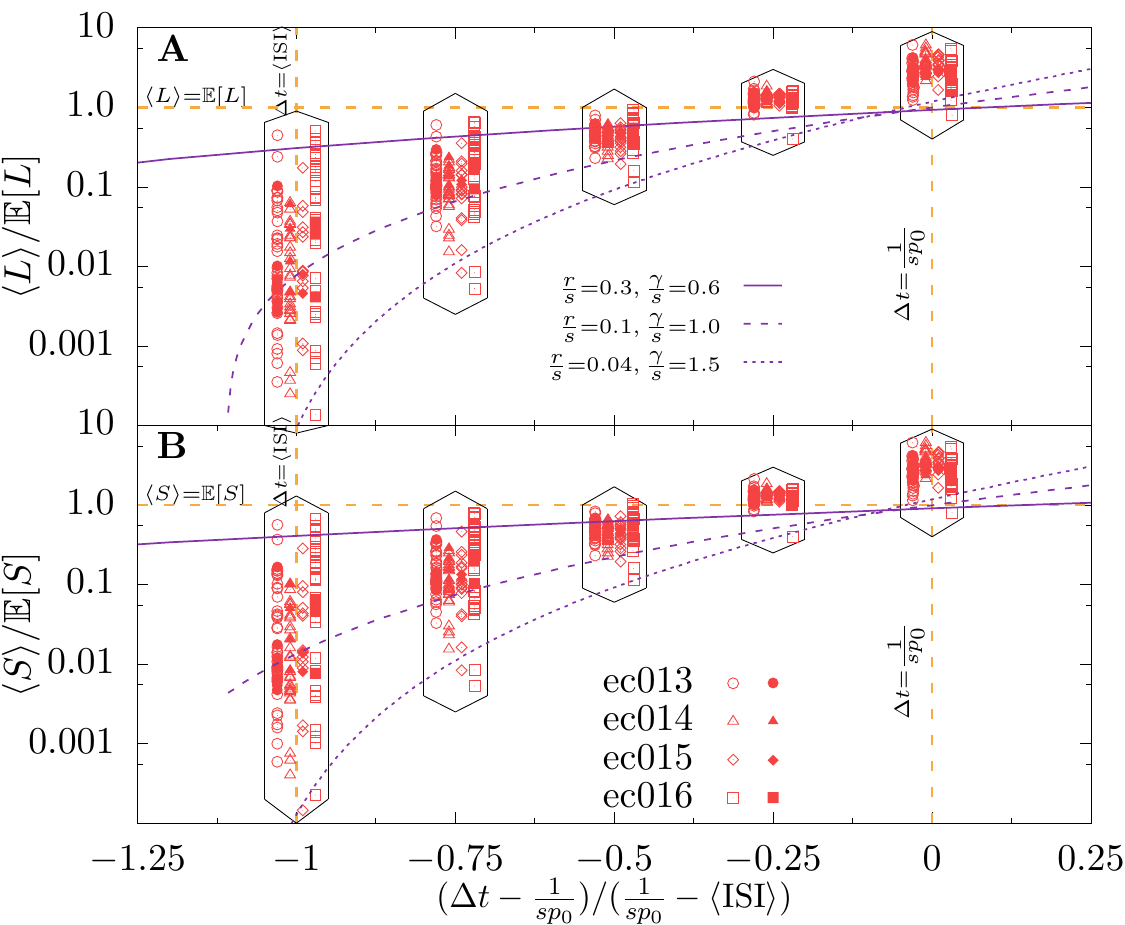}
\caption{Accuracy of avalanche duration (Panel \textbf{A}) and size (Panel \textbf{B}) for varying time bins $\Delta t$. The $x$ axis is scaled such that the $\Delta t=\langle\text{ISI}\rangle$ is at $x=-1$ and the expected single particle extinction time $\Delta t=\frac{1}{sp_0}$ is at $x=0$ for all choices of $\frac{r}{s}$, $\frac{\gamma}{s}$ and $s$. The $y$ axis is scaled by the theoretically predicted duration or size, such that $y=1$ corresponds to equality between true value and the value retrieved when using time bins. Purple lines: simulations of pumped branching processes resampled using time bins $\Delta t$. Symbols: \textit{in vivo} data from hc-2 from crcns.org.\cite{Buzsaki2009}. Solid symbols are data from combined time series of all shanks of the probe, hollow symbols are data from individual shanks. Each framed group of data points belongs to either $y\in\{-1,-0.75,-0.5,-0.25,0\}$, they are slightly shifted left and right of these values to distinguish between data sets from different rats. Dashed orange lines are visual guides.}
\label{fig-avalanche-size-length-time-bin}
\end{center}
\end{figure}

Now that all the parameters are determined, avalanche durations and sizes can be considered. Attaching time bins $\Delta t$ to spikes generates times series of avalanches. The choice of $\Delta t$ changes the avalanche properties \cite{Priesemann2009}. On the one hand, these changes can be determined for the model using simulations, and on the other hand, they can be found for the available data in the usual way. The former are created by simulating continuous-time pumped branching processes and attaching time bins $\Delta t$ to every creation/spike event, resulting in a new branching-like process of which avalanche duration and size distributions can be determined numerically. The results of these simulations are shown for example parameter choices in Fig.~\ref{fig-avalanche-size-length-time-bin} (purple lines).  In Fig.~\ref{fig-avalanche-size-length-time-bin}, the $x$ axis is scaled and shifted such that $\Delta t=\langle\text{ISI}\rangle$ is at $x=-1$ and $\Delta t=\frac{1}{sp_0}$ is at $x=0$ for all parameter choices of $\frac{r}{s}$ and $\frac{\gamma}{s}$. On the $y$ axis, the observed duration $\langle L\rangle$ or size $\langle S\rangle$ is scaled by the true expected duration $\mathbb{E}[L]$ or size $\mathbb{E}[S]$ of the pumped branching process, such that $y=1$ if they are equal for all parameter choices. It shows that duration and size are significantly underestimated at $\Delta t=\langle\text{ISI}\rangle$ ($x=-1$) and are only slightly overestimated at $\Delta t=\frac{1}{sp_0}$ ($x=0$). The latter implies that a mean-field approach in time, i.e. fixing $\Delta t$ to the expected extinction time also causes errors, although smaller ones. This analysis can also be applied to the experimental data. Here we only show \textit{in vivo} data as red symbols in the figure because the STS assumption is clearly violated for this data. Although slightly spread for better visibility, the data points lie only on values $x\in\{-1,-0.75,-0.5,-0.25,0\}$. Fig.~\ref{fig-avalanche-size-length-time-bin} shows that the \textit{in vivo} experimental data follows the same behavior for avalanche duration and size as theoretically predicted with slightly larger errors at $\Delta t=\frac{1}{sp_0}$ ($x=0$).  

\begin{figure}
\begin{center}
\includegraphics[width=\columnwidth]{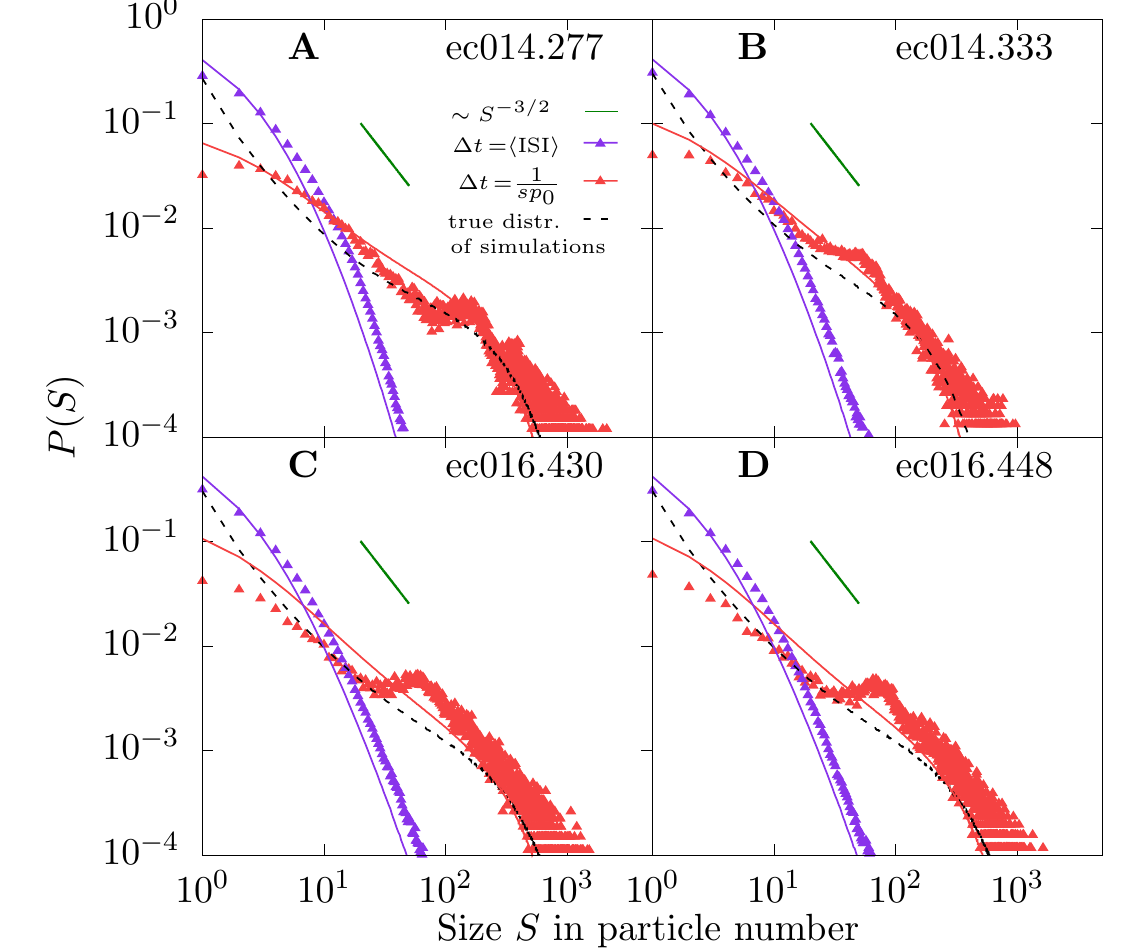}
\caption{Avalanche size distributions for four example data sets.  Symbols: \textit{in vivo} experimental data \cite{Buzsaki2009} with $\Delta t=\langle\text{ISI}\rangle$ (purple) and $\Delta t=\frac{1}{sp_0}$ (red). Solid line: simulation of avalanche size distribution of pumped branching process with matched parameters as data set and resampled with same bin sizes, $\langle\text{ISI}\rangle$ (purple) and $\frac{1}{sp_0}$ (red). Black dashed lines: true avalanche size distributions of the pumped branching process with exponentially distributed single particle extinction time according to parameters $\frac{r}{s}$, $\frac{\gamma}{s}$ and $s$ of data set. The local maximum in experimental data could be a finite size effect of the probe. Green line: visual guide for a power law of $S^{-3/2}$.}
\label{fig-avalanche-size-distribution}
\end{center}
\end{figure}

Of particular interest has been the avalanche size distribution.\cite{Beggs2003,Beggs2004,Priesemann2009} Fig.~\ref{fig-avalanche-size-distribution} shows avalanche size distributions for four \textit{in vivo} example data sets. The shown combined data sets were chosen because they had the most avalanches within the recorded data, thus enabling better statistics. Each panel shows the size distribution using $\Delta t=\langle\text{ISI}\rangle$ (purple) and $\Delta t=\frac{1}{sp_0}$ (red) of the data (symbols) and simulations (same colored lines) of the process with matched parameters and bin size. In addition, each plot also shows as black dashed line the true theoretical avalanche size distribution, i.e. without time binning. Firstly, for $\Delta t=\langle\text{ISI}\rangle$ (purple), experimental data and simulations agree very well. For $\Delta t=\frac{1}{sp_0}$ (red), experiment and simulation agree on the general trend, however the experimental data has a local maximum around the size of 100 particles, which might be explained as a finite size effect of the probe. The true theoretical size distribution agrees better with the time bin choice $\Delta t=\frac{1}{sp_0}$, but shows considerable differences for small avalanche sizes. This might be explained by considering that short avalanches with one or two particles might be attached to longer ones due to the binning procedure. Overall, the model provides a fairly accurate prediction of the influence of binning on the data's avalanche distributions. In particular, this result might imply that \textit{in vivo} neuronal activity is \textit{not scale free} and is \textit{not at criticality} for the used experimental data set.\cite{Buzsaki2009} 

\subsection{Spatial subsampling}\label{sec-subsampling}
\begin{figure}
\begin{center}
\includegraphics[width=\columnwidth]{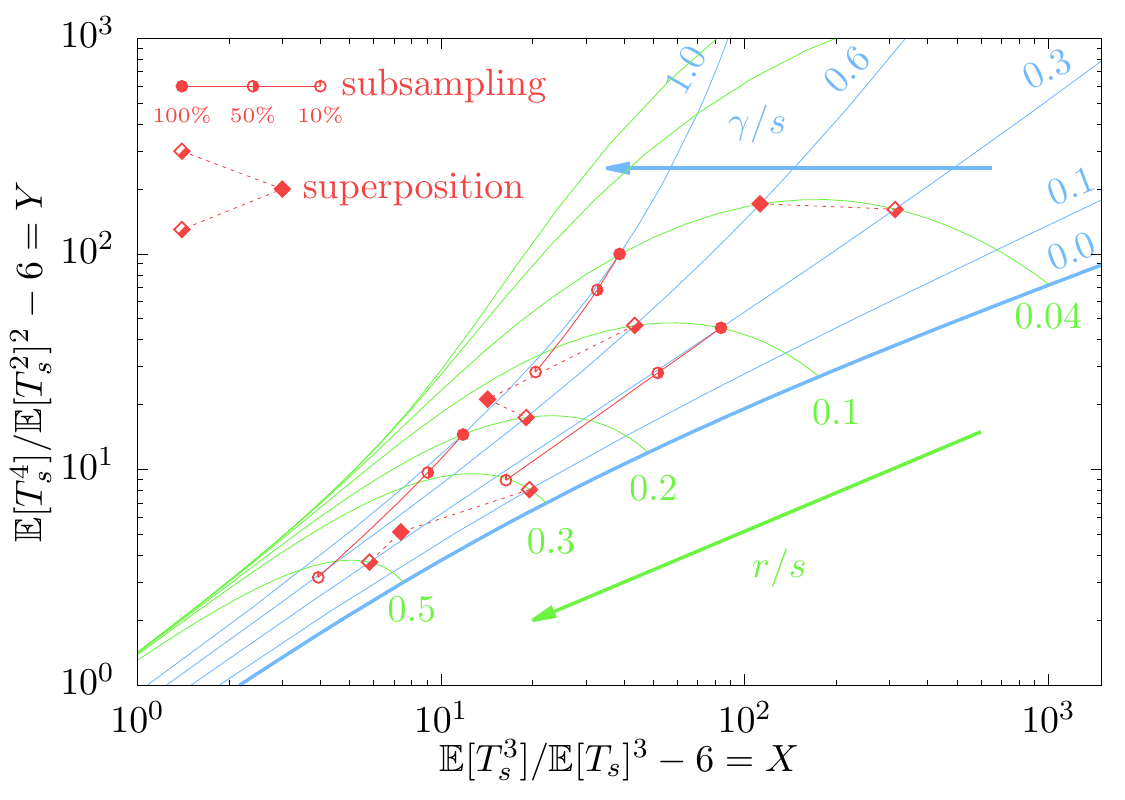}
\caption{Moment-ratio map with mean-field subsampling. Blue lines are $\frac{\gamma}{s}$-level sets, green lines are $\frac{r}{s}$-level sets. Circle symbols: random subsampling of a process to 50\% and 10\% of spikes results in decrease of apparent relative spontaneous creation $\frac{\gamma}{s}$ and degree of criticality $\frac{r}{s}$. Diamond symbols: overlapping two independent processes leads to an apparent  mix of the individual degrees of criticality $\frac{r}{s}$ and an increase in relative spontaneous creation $\frac{\gamma}{s}$. The experimental data did not clearly fit into these patterns, implying that the subsampling is beyond-mean-field behavior.}
\label{fig_subsampling}
\end{center}
\end{figure}
Subsampling is a common problem in experiments and a difficult problem when neuronal spikes are measured.\cite{Beggs2003,Beggs2008,Priesemann2009,Levina2017,Wilting2018b,Zierenberg2020} In Sec.~\ref{sec-twotimecorrelation}, it was discussed what effect temporal subsampling can have on estimating the degree of criticality from the auto-correlation function. In Sec.~\ref{sec-ISI-extinction-time}, it was shown what effect temporal binning, i.e. a forced corse temporal sampling, can have on avalanche statistics.  

However, \textit{spatial} subsampling is also an issue when measuring neuronal spiking. Estimating its influence involves assumptions about the nature of the subsampling. One possible assumption is mean-field subsampling, i.e. assuming that the observed sample is an unknown and randomly chosen proportion of the entire spiking activity. This can be assumed for observations of the entire spatial extension of the system (such as in disease control). But it is also possible in compartmentalized systems, if it is assumed that one compartment is an accurate representation of all compartments (such as in quality control). Mean-field subsampling assumes that one sample does not affect another sample. This mean-field-type subsampling was studied in \cite{Levina2017,Zierenberg2020} using mostly simulations and analytics.  In \cite{Zierenberg2020}, the  `expected rate estimator' approach is closest to the zero-dimensional approach presented here. In the relevant subcritial regime, \cite{Zierenberg2020} showed good agreement with the microscopic network parameters, see in particular Fig.~4 in \cite{Zierenberg2020}, which could be regarded as a justification in this article to ignore spatial components. However, the assumptions in \cite{Levina2017,Zierenberg2020} of 1) mean-field subsampling, 2) minor influence of time-binning and 3) separation of time scales should be investigated, in particular since assumptions 2) and 3) have been found to be invalid for the used \textit{in vivo} data in Sec.~\ref{sec-ISI-extinction-time} above. In the following, the assumption of mean-field subsampling is checked by simulating its effects and comparing it to results of subsampling of the used experimental \textit{in vivo} data.\cite{Buzsaki2009}  

For the spike time series induced by pumped branching processes, the simulated effect of mean-field sampling is shown in Fig.~\ref{fig_subsampling} with circle symbols. It shows the moment ratios $X$ and $Y$, Eq.~\eqref{eq-moment-ratios}, of the original process (filled circle), subsampled at 50\% of the activity (half circle) and 10\% (hollow circle). The $\frac{r}{s}$ and $\frac{\gamma}{s}$ level sets in the figure show that mean-field subsampling results in reduced degree of criticality $\frac{r}{s}$ and reduced  relative spontaneous  creation $\frac{\gamma}{s}$.

While it is clear that electrode arrays only measure a small fraction of the brain, it is unclear whether they actually subsample a single spiking process. It could be that the different shanks of the probe or different areas of the electrode array sample different independent spiking processes simultaneously.  Fig.~\ref{fig_subsampling} also shows the expected effect of this phenomenon with diamond symbols. Half-filled diamonds represent independent spike processes, while filled diamonds indicate the moment ratios of the combined process (i.e. overlapped, not concatenated). In this case, the result is a mixed degree of criticality and an increased relative spontaneous creation.

In both scenarios (re)combining data (solid symbols) leads to increased relative spontaneous creations $\frac{\gamma}{s}$ compared to the individual subprocesses. This is expected, as spontaneous creation is an independent subprocess that does not depend on the present particle number in the system.  

The experimental \textit{in vivo} data \cite{Buzsaki2009} in Fig.~\ref{fig_SpikeMomentsMapExpData} includes the data of individual shanks (hollow symbols) and the combined data of all shanks of a probe (solid symbols). It indicates that the combined experimental data has on average \textit{reduced} relative spontaneous creation rates $\frac{\gamma}{s}$ and slightly higher degree of criticality $\frac{r}{s}$ (lower $\frac{r}{s}$), which can be seen better in Fig.~\ref{fig_SpikeMomentsMapExpData}\textbf{B}. This reduction cannot be explained by mean-field subsampling. Does this mean the underlying model is insufficient or wrong? Not necessarily. The shifts seen in the combined data can be explained by \textit{beyond-mean-field} subsampling, where samples are neither independent nor a random fraction of the complete activity. Beyond-mean-field subsampling depends on the connectome, i.e. the precise map of neuronal connections. It can occur as follows: In sampled area \textit{A} it might appear that a neuron does not propagate the signal to other neurons, i.e. it inhibits the signal, because the receiving neurons lie outside the locally sampled area. Thus, area \textit{A} appears to be further away from criticality. Conversely, in area \textit{B} a neuron might appear to be spontaneously activated because its signal source lies outside of area \textit{B}. Thus, area \textit{B} appears to have a high relative spontaneous creation $\frac{\gamma}{s}$. However, if \textit{A} and \textit{B} are connected such that the receiving neuron from \textit{A} lies in \textit{B}, then the apparent high inhibition in \textit{A} is actually explained by the apparent strong spontaneous creation in \textit{B}. In terms of parameters, this means that a larger $\frac{r}{s}$ in \textit{A} and a larger $\frac{\gamma}{s}$ in \textit{B} are resolved in a lower $\frac{r}{s}$ (i.e. higher degree of criticality) and a lower $\frac{\gamma}{s}$ in the combined data of \textit{A} and \textit{B}. This phenomenon is visible to some extent in Fig.~\ref{fig_SpikeMomentsMapExpData}\textbf{B}. Combined data (solid symbols) appears to be identified on average with lower $\frac{\gamma}{s}$ and $\frac{r}{s}$ than single-shank data (hollow symbols).  

While these observables are time scale independent, the estimated time scale $s$ is significantly changed in the combined \textit{in vivo} data: Fig.~\ref{fig-ISI-vs-extinction-time} shows a clear trend of smaller $\langle\text{ISI}\rangle$ and expected extinction time $\frac{1}{sp_0}$ for combined \textit{in vivo} data compared to single-shank data (red symbols). 

\section{Conclusion}\label{sec-conclusion}
This article presents a novel approach for determining how close neuronal avalanches are to criticality. Instead of generating a branching-like process, the approach is an analytical derivation of spike time series from pumped branching processes. In detail: the new method does not require processing spike-time recordings by attaching time bins and analysing the resulting step process. The new method directly works with spike time recordings and calculates ratios of the moments of the inter-spike interval. Once the ratios are calculated, the parameters of the process can be read off a map, the moment-ratio map, Fig.~\ref{fig_SpikeMomentsMapExpData}. The user can generate this map to the desired precision by using the provided python code\cite{Pausch2020b}.  Some conceptual drawbacks were pointed out in the setup of the model, Sec.~\ref{sec-model}. The new approach has some significant success, of which several are highlighted in the following. 

On the bare level of inter-spike intervals, the new method predicts moments, the coefficient of variation and its skewness to high accuracy. Their approximation errors were analysed in Fig.~\ref{fig_Approximation_Error},  which showed small percentage errors. This shows that the method and process represent the observed spike time series very well. 

In comparison to other methods, the approach avoids creating a branching-like processes and allows analysing inter-spike intervals directly. In particular, the proposed methodology does not require time-binning and the defining parameters (degree of criticality $\frac{r}{s}$ and relative spontaneous creation $\frac{\gamma}{s}$) are dimensionless. Therefore, it is less sensitive compared to a number of commonly used data processing methods.\cite{Priesemann2009,Wilting2018b}  More precisely, it allows identifying how close the system is to criticality with a time-scale independent measure and can therefore be regarded advantageous over criticality estimates based  on either time-bin-dependent power-law distributions of avalanche sizes or on time-bin-dependent auto-correlation functions.\cite{Priesemann2009,Wilting2018b} In particular, the proposed new method applied to experimental data indicates that neuronal circuits are further away from criticality than most articles suggest.\cite{Beggs2003,Beggs2004,Beggs2008,Priesemann2009,Priesemann2013,Wilting2018b,Wilting2018a} 

The most common way of confirming criticality in the brain has been the identification of a power law distribution of avalanche sizes. However, that approach is a notoriously difficult one in statistical physics and is prone to introduce biases.\cite{Goldstein2004,Priesemann2018} These difficulties are avoided in this article by looking at a moment-ratio map of the inter-spike intervals. It avoids generating a branching-like time series and thus also avoids fitting power laws. In addition, the effects of binning on a pumped branching process strongly resemble its observed effects when applied to experimental data, Figs.~\ref{fig-avalanche-size-length-time-bin} and~\ref{fig-avalanche-size-distribution}. Consequently and more importantly, it indicates that neuronal avalanche dynamics in the used \textit{in vivo} data sets \cite{Buzsaki2009} is not scale free and not at criticality. The used \textit{in vitro} data sets \cite{Wagenaar2006} show that criticality increases for a plated developing neural network over time and that external input is significantly lower compared to the \textit{in vivo} data. 

The external signal immigration (\textit{`pumping'}) into the locally measured neuronal circuit is also determined in this article. It suggests that for the \textit{in vivo} data, avalanche overlap is relatively common and cannot be ignored. In particular, the common separation-of-timescales (STS) assumption is not justified for the used \textit{in vivo} data \cite{Buzsaki2009}. It is therefore also \textit{expected} that the avalanche size distribution does not follow a power law distribution with exponent $-3/2$ even if it were at criticality.\cite{Munoz2017} In contrast for the \textit{in vitro} data \cite{Wagenaar2006}, external input was rare and the STS assumption is valid.

Another advantage of the presented model is that relaxation behavior can be predicted analytically. The presented methodology would allow to investigate whether different steady states,  such as wakefulness or deep sleep \cite{Priesemann2013}, relate to changes in external input $\frac{\gamma}{s}$ or to internal adjustment of branching efficiency $\frac{r}{s}$.

The presented model and data analysis also opens several routes for future research. Including topological constraints in a pumped branching process is a major challenge which will most likely have significant impact on more detailed comparisons with experiments. In addition, when considering subsampling, Sec.~\ref{sec-subsampling}, the data indicated beyond-mean-field subsampling which would require detailed knowledge of the connectome for more accurate models. Furthermore, better understanding finite size effects, which are unintentionally introduced due to the size of used probes, might clarify some of the anomalous statistics observed. In a different direction, future work could also try to understand mechanism that cause inter-avalanche correlations.

\section*{Acknowledgements}
The author wishes to thank Gunnar Pruessner, Ignacio Bordeu Weldt, Rosalba Garcia Milan, Benjamin Walter, and Viola Priesemann for helpful discussions. He received financial support through an EPSRC PhD scholarship and an EPSRC Doctoral Prize Fellowship at Imperial College London.

\appendix
\section{Derivation of spontaneous creation action}
\label{appendix-spontaneous-creation-derivation}
Let $P(N(t)=n)$ denote the probability that there are $n$ particles in the system at time $t$. Then, a spontaneous creation of particles can be described by a master equation \cite{vanKampen1992} as follows\begin{align}
\frac{\plaind}{\plaind t}P(N(t)=n)=\gamma\Bigl(P\bigl(N(t)=n-1\bigr)-P\bigl(N(t)=n\bigr)\Bigr).\label{eq-appendix-master-equation}
\end{align}
Such a master equation can be transformed into an action of a Doi-Peliti field theory \cite{Taeuber2014,Garcia-Millan2018,Pausch2019b}, which results in the action $\mathcal{A}_c$ shown in Eq.~\eqref{eq-spontaneous-creation-action}.

For comparison, the branching/extinction process with offspring distribution $p_k$ and occurrence rate $s$ is described the following master equation\begin{align}
\frac{\plaind}{\plaind t}P(N(t)=n)=&s\sum\limits_{k=0}^\infty p_k(n-k+1)P(N(t)=n-k+1)\notag\\&-snP(N(t)=n)\label{eq-master-eq-branching},
\end{align} 
derived in \cite{Garcia-Millan2018}. 

Although written as a single equation, master equations~\eqref{eq-appendix-master-equation} and~\eqref{eq-master-eq-branching} represent one ordinary differential equations for each $n\in\mathbb{N}_0$. Thus, they actually describe an infinite system of coupled ordinary differential equations. 

\section{$\langle\phi^m\rangle=\mathbb{E}[(N(t))_m]$}\label{sec-factorial-moment-outline}
In Doi-Peliti field theories, the $m$th factorial moment naturally appears as observable and takes the form $\phi^m$. This is explained at length in \cite{Pausch2019b} and a brief outline is given here. 

The actions of Doi-Peliti field theories are derived from Master equations which take the form \begin{align}
\frac{\plaind}{\plaind t} P(N(t))=\mathcal{L}P(N(t)),\label{eq-master-equation-general-form}
\end{align}
where $P(N(t))=(P(N(t)=0),P(N(t)=1),\dots)^T$ is a vector and $\mathcal{L}$ is a linear transformation. Eq.~\eqref{eq-master-equation-general-form} is a system of infinitely many coupled ordinary differential equations. It can be transformed into a single partial differential equation for the probability generating function, defined as \begin{align}\mathcal{M}(z,t)=\sum\limits_{n=0}^\infty P(N(t)=n)z^n,
\end{align}
by using derivatives w.r.t. $z$ or multiplying by $z$. For example, Eq.~\eqref{eq-appendix-master-equation} is transformed into\begin{align}
\frac{\plaind}{\plaind t}\mathcal{M}(z,t)=\gamma\left(z-1\right)\mathcal{M}(z,t),
\end{align}
and Eq.~\eqref{eq-master-eq-branching} is transformed into\begin{align}
\frac{\plaind}{\plaind t}\mathcal{M}(z,t)=s\left(\sum\limits_{n=0}^\infty p_kz^{k} -z\right)\frac{\plaind}{\plaind z}\mathcal{M}(z,t).
\end{align}
From this PDE perspective, path integrals are just solving PDEs perturbatively around $z=0$.
However in most references, $z$ and $\plaind/\plaind z$ are represented as ladder operators $a^\dagger$ and $a$, respectively, and interpreted as particle creators and annihilators.

A useful step is introducing a new variable $\widetilde z$ by setting $z=\widetilde z+1$, and work with $\widetilde{\mathcal{M}}(\widetilde z,t)=\mathcal{M}(z,t)$.  This is called the Doi-shift. The path integral solves now the corresponding PDE for $\widetilde{\mathcal{M}}$ perturbatively around $\widetilde z=0$, i.e. \begin{align}
\langle\phi^m(t)\rangle=\frac{\plaind^m}{\plaind z^m}\widetilde{\mathcal{M}}(\widetilde z,t)\bigr|_{\widetilde z=0}=\mathbb{E}[(N(t))_m],
\end{align}
where $(N)_m=N(N-m)\dots(N-m+1)$ is the falling factorial.

\section{Steady-state probability distribution}\label{appendix-steadystateprobability}
Using the fact that the probability generating function is equal to the factorial moment generating function, the probability distribution of the particle number $N$ can be deduced from the factorial moments $\langle\phi^k\rangle$:
\begin{align}
P(N=\ell)=&\frac{1}{\ell!}\frac{\plaind^\ell}{\plaind z^\ell}M_N(z)\Biggr|_{z=0}=\sum\limits_{k=\ell}^\infty\frac{(-1)^{k-\ell}}{(k-\ell)!}\left(\frac{q_2}{r}\right)^k\left(\frac{\gamma}{q_2}\right)^{(k)}\notag\\
=&\frac{1}{\ell!}\left(\frac{q_2}{r}\right)^\ell\sum\limits_{m=0}^\infty\frac{(-1)^{m}}{m!}\left(\frac{q_2}{r}\right)^m\left(\frac{\gamma}{q_2}\right)^{(m+\ell)}\notag\\
=&\frac{1}{\ell!}\left(\frac{q_2}{r}\right)^\ell\left(\frac{\gamma}{q_2}\right)^{(\ell)}\sum\limits_{m=0}^\infty\frac{(-1)^{m}}{m!}\left(\frac{q_2}{r}\right)^m\left(\frac{\gamma}{q_2}+\ell\right)^{(m)}\notag\\
=&\frac{1}{\ell!}\left(\frac{q_2}{r}\right)^\ell\left(\frac{\gamma}{q_2}\right)^{(\ell)}\sum\limits_{m=0}^\infty\frac{1}{m!}\left(\frac{q_2}{r}\right)^m\left(-\frac{\gamma}{q_2}-\ell\right)_{m}\notag\\
=&\frac{1}{\ell!}\left(\frac{q_2}{r}\right)^\ell\left(\frac{\gamma}{q_2}\right)^{(\ell)}\left(1+\frac{q_2}{r}\right)^{-\frac{\gamma}{q_2}-\ell}
\end{align}
where in the second to last line, the umbral Taylor expansion of $(1+q_2/r)^{-\gamma/q_2-\ell}$ around $q_2/r=0$  is identified. $(x)^{(\ell)}=x(x+1)\dots(x+\ell-1)$ is the rising factorial.

Let's check that this probability distribution is normalized:\begin{align}
\sum\limits_{\ell=0}^\infty P(N=\ell)=&\frac{1}{\left(1+\frac{q_2}{r}\right)^\frac{\gamma}{q_2}}\sum\limits_{\ell=0}^\infty\frac{1}{\ell!}\left(\frac{q_2}{r+q_2}\right)^\ell\left(\frac{\gamma}{q_2}\right)^{(\ell)}\notag\\
=&\frac{1}{\left(1+\frac{q_2}{r}\right)^\frac{\gamma}{q_2}}\sum\limits_{\ell=0}^\infty\frac{1}{\ell!}\left(\frac{-q_2}{r+q_2}\right)^\ell\left(\frac{-\gamma}{q_2}\right)_{\ell}\notag\\
=&\frac{1}{\left(1+\frac{q_2}{r}\right)^\frac{\gamma}{q_2}}\left(1-\frac{q_2}{r+q_2}\right)^{-\frac{\gamma}{q_2}}=1
\end{align}

\section{Initializing an empty system}\label{sec-appendix-empty-systen}
In the operator picture, where $z=a^\dagger$ and $\frac{\plaind}{\plaind z}=a$, states of the system are described as Fock states $|n\rangle=z^n$. Their dual states over the $L^2$ inner product are compactly written as $\langle n|$. Before time $t_0=0$, the system is allowed to evolve freely resulting in state $|\mathcal{M}(t_0^-)\rangle=\sum_{n=0}^\infty P(N(t_0^-)=n)|n\rangle$. But at $t_0=0$, it is projected out using $\langle\sun|=\sum_{n=0}^\infty\langle n|$. Because $\langle m|n\rangle=\delta_{mn}$, the result is $\langle\sun|\mathcal{M}(t_0^-)\rangle=1$. Then, the system is reinitialized with zero particles by attaching the empty state $|0\rangle$. Thus, given that the system was empty at time $t_0=0$,  the $n$th factorial moment at time $t$ equals\begin{align}
\mathbb{E}[(N)_n(t)|N(0)=0]=\langle\sun|a^{\dagger n}a^ne^{\mathcal{A}t}|0\rangle\langle\sun|\mathcal{M}(t_0)\rangle
\end{align}
In the field theory, the projection with $\langle\sun|$ and reinitialization with $|0\rangle$ is expressed as the observable $e^{-\widetilde\phi(0)\phi(0)}$. See \cite{Pausch2019b} for more details on how to derive the field-theoretic expressions from operator expressions.

\section{Inter-event time distribution}\label{sec-event-time-distribution}
In steady state, the branching process' particle number probability distribution is a negative binomial (or P\'olya) distribution with the general form\begin{align}
P(N=\ell)=\frac{1}{\ell!}\frac{\Gamma(u+\ell)}{\Gamma(u)}p^u(1-p)^\ell,
\end{align}
with $u=\gamma/q_2$, $p=r/(r+q_2)$. Its moment generating function equals\begin{align}
\mathcal{M}_N(x)=\left(\frac{p}{1-(1-p)e^x}\right)^u\qquad\text{for }x<-\ln(1-p)
\end{align}
This distribution can be regarded as a prior distribution to a conditional exponential distribution $f_e(t|\ell)$ with rate $\lambda=\ell s+\gamma$:\begin{align}
f_e(t|\ell)=(s\ell+\gamma)e^{-(s\ell+\gamma)t}.
\end{align}
The marginal event time distribution $f_e(t)$ can be recovered as \begin{align}
f_e(t)=&\sum\limits_{\ell=0}^\infty f_e(t|\ell)P(N=\ell)\\
=&e^{-\gamma t}\left(\gamma-\frac{\partial}{\partial t}\right)\mathcal{M}_N(-st)\notag\\
=&\gamma e^{-\gamma t}\left(\frac{(r+q_2)(1+e^{-st})}{r+q_2(1-e^{-st})}\right)\left(\frac{r}{r+q_2(1-e^{-st})}
\right)^{\frac{\gamma}{q_2}}\notag
\end{align}

\section{Verification of moments of the inter-event and inter-spike interval}\label{sec-interval-verification}
\begin{figure}
\begin{center}
\includegraphics[width=\columnwidth]{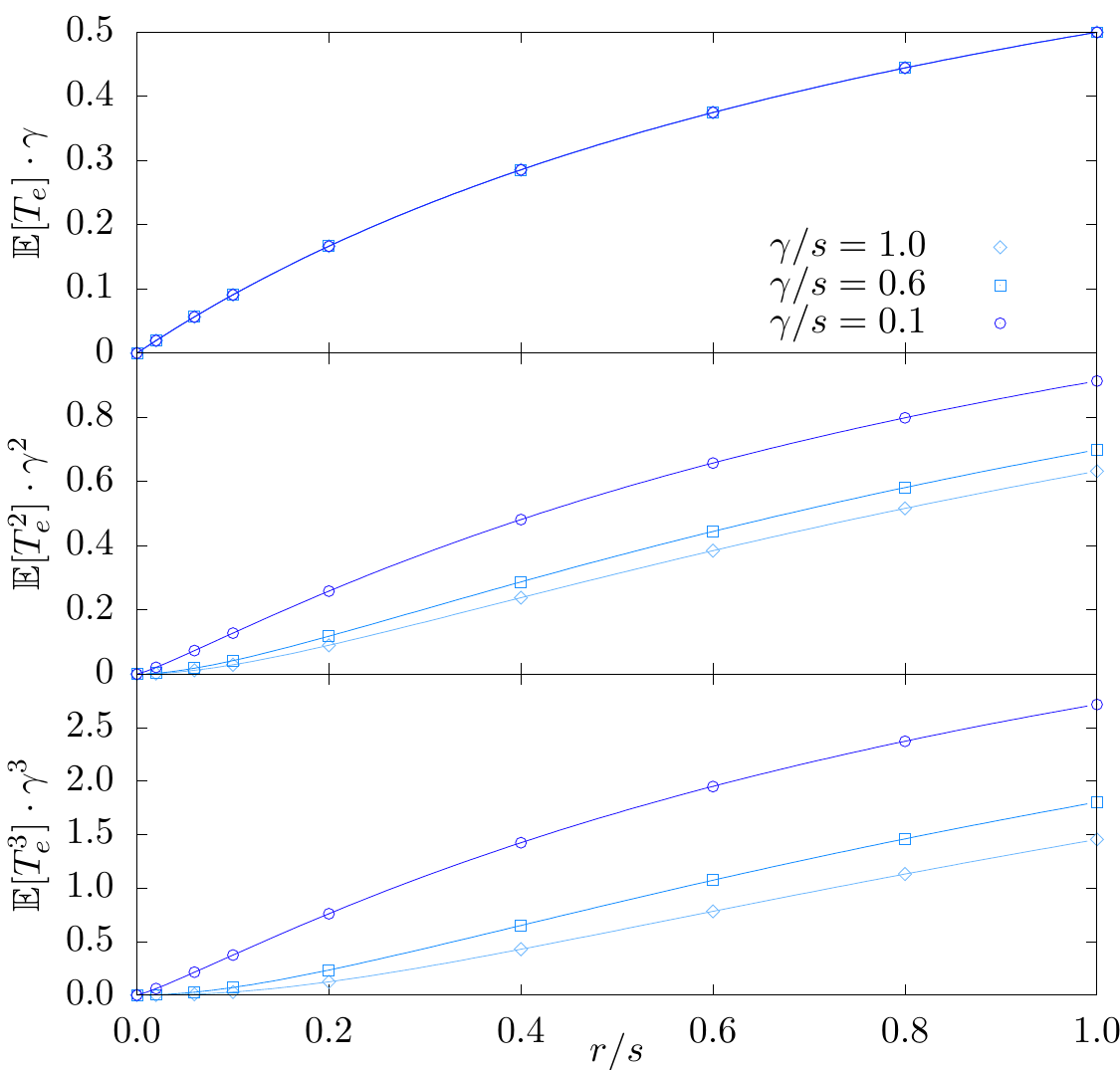}
\end{center}
\caption{Rescaled first, second and third moment of the time between events over $r/s\in[0,1]$. Events include both extinctions and creations. The parameter values are $\gamma/s\in\{0.1,0.6,1.0\}$. Symbols: simulation results. Lines: Analytical prediction. For the first moment, all lines collapse after rescaling.}\label{fig-event-time-moments}
\end{figure}

It is important to verify analytical derivations as much as possible and reasonable and to make such verifications available. Here, the moments of the inter-event time and the inter-spike time are verified using  simulations in the parameter space that is relevant for comparison with experimental data. Fig.~\ref{fig-event-time-moments} shows the first three moments of the inter-event interval over a range of $\frac{r}{s}$, the degree of criticality. As expected, intervals become shorter as criticality $\frac{r}{s}=0$ is approached. 

\begin{figure}
\begin{center}
\includegraphics[width=\columnwidth]{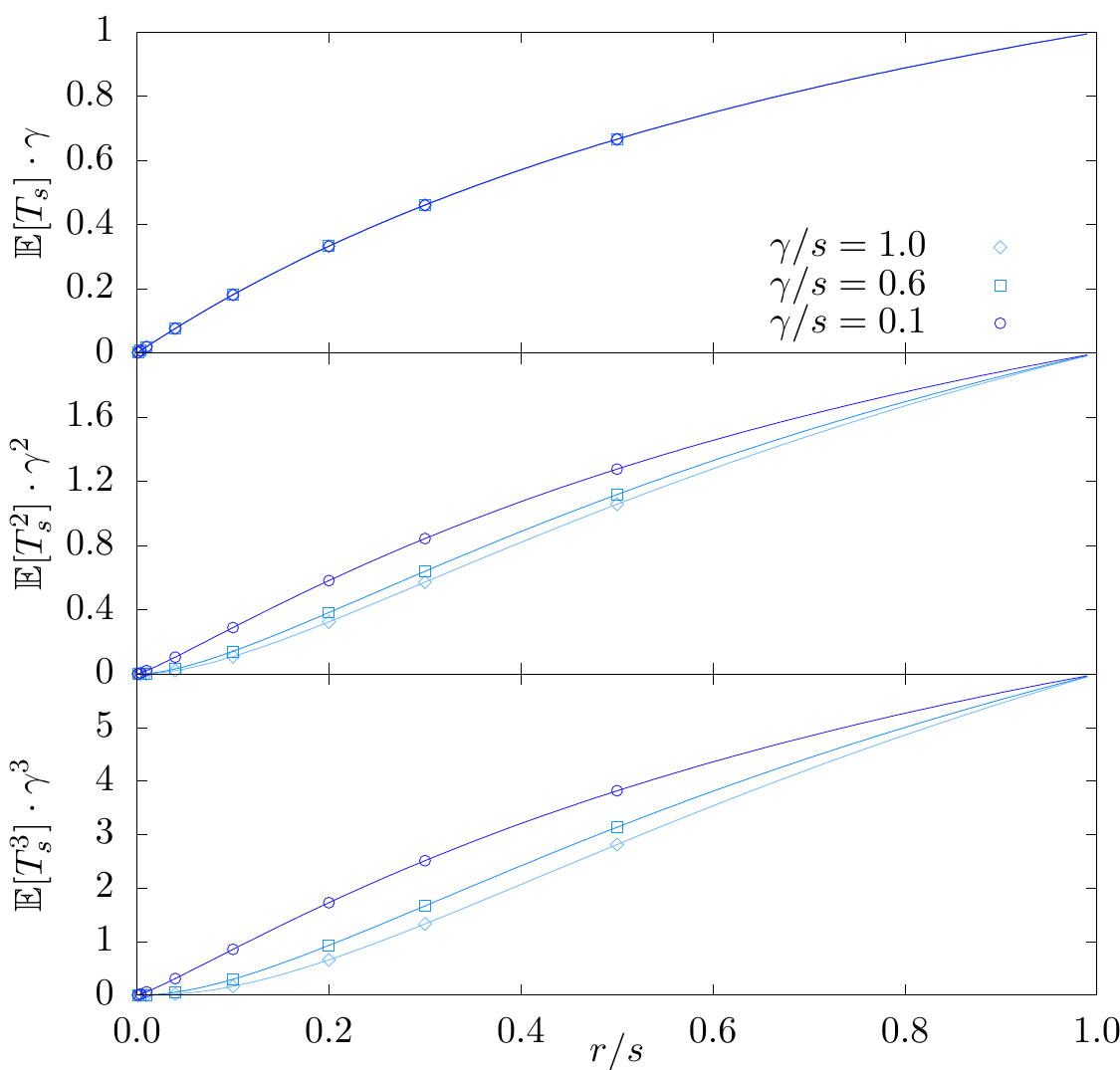}
\end{center}
\caption{Rescaled first, second and third moment of the time between spikes, i.e. between creations of particles, over $r/s\in[0,1]$. The parameter values are $\gamma/s\in\{0.1,0.6,1.0\}$. Symbols: simulation results. Lines: Analytical prediction. For the first moment, all lines collapse after rescaling.}\label{fig-spike-time-moments}
\end{figure}

Fig.~\ref{fig-spike-time-moments} shows the first three moments of the inter-spike interval over a range of $\frac{r}{s}$ values. Analogously to the inter-event intervals, the inter-spike intervals tend to zero as criticality is approached. 

Although the moments of intervals between events and spikes appear to be very similar, they show an important difference in the limit of a pure Poisson process $\frac{r}{s}\rightarrow1$. In this limit, spike times will be equal to spontaneous creation times which occur with rate $\gamma$. Event intervals however still depend on the extinction time scale $sp_0\rightarrow s$.

\section{Derivation of the moments of inter-spike intervals given initial state}
\label{appendix-spike-waiting-time-from-fixed-state}
The result in Eq.~\eqref{eq-spike-waiting-time-from-fixed-state} can be found by starting to calculate it explicitly for $m=0,1,2,\dots$ from which the general rule can be derived.

Given the system is in state $n=0$, the next creation event time is $\sim\text{Exp}(\gamma)$ distributed. If the system is in state $n=1$, the next creation event time has the following moments\begin{align}
\mathbb{E}[T_c^n|N=1]=&\int\limits_{0}^\infty \biggl(t^n(s+\gamma) e^{-(s+\gamma)t}\frac{sp_2+\gamma}{s+\gamma}+\\
&\hspace{-2cm}+\int\limits_{0}^\infty (t+t')^n(s+\gamma) e^{-(s+\gamma)t}\frac{sp_0}{s+\gamma}\gamma e^{-\gamma t'}\plaind t'\biggr)\plaind t\notag\\
&\hspace{-2.3cm}=(sp_2+\gamma)\left(-\frac{\plaind}{\plaind \gamma}\right)^n\frac{1}{s+\gamma}+sp_0\gamma\left(-\frac{\plaind}{\plaind \gamma}\right)^n\frac{1}{\gamma(s+\gamma)}.\notag
\end{align}
If the system is in state $n=2$, the moments are\begin{align}
\mathbb{E}[T^n|N=2]=&\int\limits_{0}^\infty t^n(2sp_2+\gamma)e^{-(2s+\gamma)t}\plaind t+\\
&\hspace{-2cm}+\int\limits_{0}^\infty \int\limits_{0}^\infty (t+t')^n2p_0 e^{-(2s+\gamma)t}(sp_2+\gamma)e^{-(s+\gamma)t'}\plaind t'\plaind t+\notag\\
&\hspace{-2cm}+\iiint\limits_{0^3}^{\infty^3} (t+t'+t'')^n 2(sp_0)^2\gamma e^{-(2s+\gamma)t-(s+\gamma)t'-\gamma t''}\plaind t''\plaind t'\plaind t\notag\\
&\hspace{-2cm}=(2sp_2+\gamma)\left(-\frac{\plaind}{\plaind\gamma}\right)^n \frac{1}{2s+\gamma}+\notag\\
&\hspace{-2cm}+2sp_0(sp_2+\gamma)\left(-\frac{\plaind}{\plaind \gamma}\right)^n\frac{1}{(s+\gamma)(2s+\gamma)}+\notag\\
&\hspace{-2cm}+2sp_0 sp_0\gamma \left(-\frac{\plaind}{\plaind \gamma}\right)^n \frac{1}{\gamma(s+\gamma)(2s+\gamma)}\notag
\end{align}
This can be generalized by induction for any condition $N=m$. If the system is in state $N=m$, then after an exponentially distributed amount of time with rate $sm+\gamma$, either a particle is created, i.e. a spike occurs, or a particle goes extinct. After the the extinction event, the previous result $\mathbb{E}[T_s^{n}|N=m-1]$ can be used and the result is shown in Eq.~\eqref{eq-spike-waiting-time-from-fixed-state}.

\section{Phase space boundary for $\gamma\rightarrow0$}\label{sec-phase-space-boundary}
Considering Eq.~\eqref{eq-spike-waiting-time-from-fixed-state}, in the limit $\gamma\rightarrow0$, the inter-spike interval moments conditioned on initial state $N=m$ diverge as follows\begin{align}\label{eq-boundary-conditioned-spike-time}
\lim\limits_{\gamma\rightarrow0}\mathbb{E}[T_s^n|N=m]\sim&\frac{n!p_0^{m}}{\gamma^n}
\end{align}

On the other hand, the limit of $f(N=m)$ for $\gamma\rightarrow0$ is\begin{align}
\lim\limits_{\gamma\rightarrow0}f(N=m)=&\,\lim\limits_{\gamma\rightarrow0}\frac{r^{u+1}\Gamma(u+m)q_2^m(sm+\gamma)}{m!\Gamma(u)\gamma(r+s)(r+q_2)^{u+m}}\notag\\
=&\,\frac{srq_2^{m-1}}{(r+s)(r+q_2)^m},
\end{align}
which implies for $f(m-1\uparrow m)$\begin{align}\label{eq-boundary-spike-event-distr}
\lim\limits_{\gamma\rightarrow0}f(m-1\uparrow m)=\frac{srq_2^{m-2}}{(r+s)(r+q_2)^{m-1}}2p_2.
\end{align}
Combining Eq.~\eqref{eq-boundary-conditioned-spike-time} and Eq.~\eqref{eq-boundary-spike-event-distr} gives\begin{align}
\lim\limits_{\gamma\rightarrow0}\mathbb{E}[T_s^n]\sim\frac{n!}{\gamma^n}\frac{2r}{(r+s)}.
\end{align}
This implies for the coefficient of variation $c_V$\begin{align}
\lim\limits_{\gamma\rightarrow0}c_V=\sqrt{\frac{s}{r}}.
\end{align}
Furthermore, the following ratios can be calculated:\begin{align}
x=&\lim\limits_{\gamma\rightarrow0}\frac{\mathbb{E}[T_s^3]}{\mathbb{E}[T_s]^3}-6=6\left(\frac{(r+s)^2}{4r^2}-1\right)\\
\Leftrightarrow&\, \frac{(r+s)}{2r}=\sqrt{(x+6)/6}\\
y=&\lim\limits_{\gamma\rightarrow0}\frac{\mathbb{E}[T_s^4]}{\mathbb{E}[T_s^2]^2}-6\\
=&\,6\left(\frac{(r+s)}{2r}-1\right)=6\left(\sqrt{(x+6)/6}-1\right),
\end{align}
which describes the phase space boundary in the moment-ratio map.

\section{Verification of the moment-ratio map}\label{sec-verification-ratio-map}
\begin{figure}
\begin{center}
\includegraphics[width=\columnwidth]{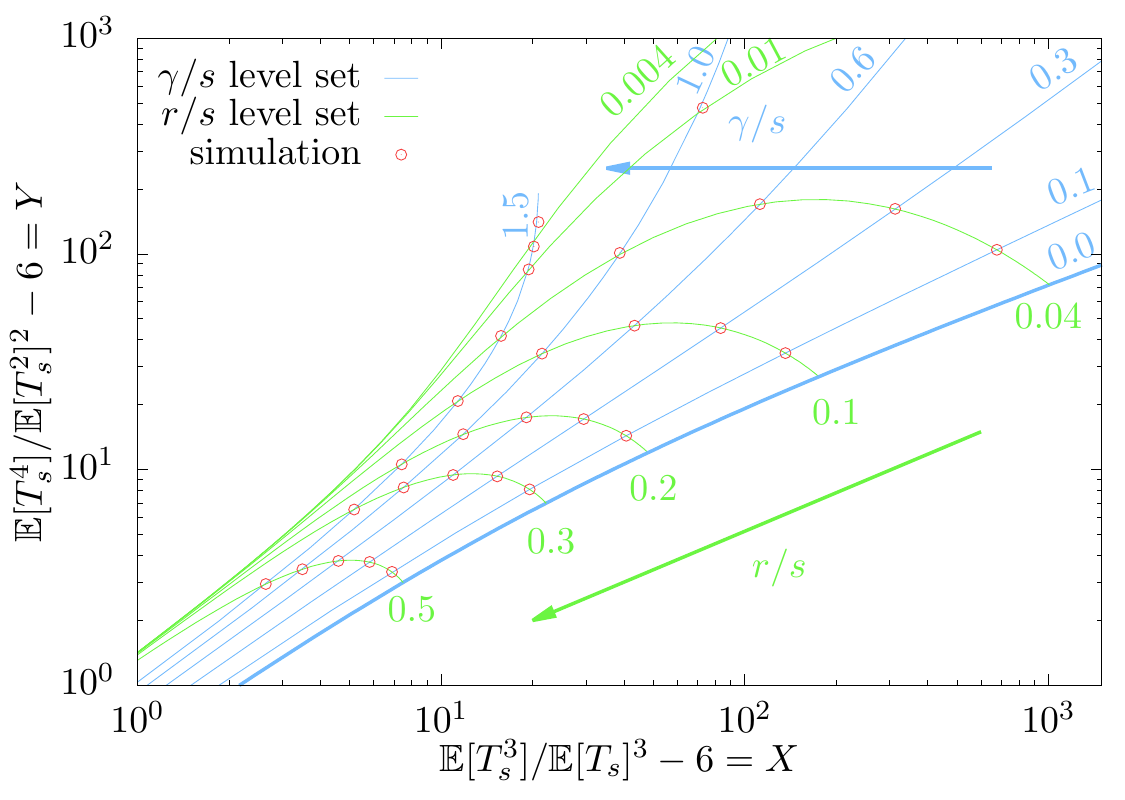}
\caption{Moment-ratio map. Blue lines: analytically calculated  $\frac{\gamma}{s}$-level sets. Green lines: analytically calculated  $\frac{r}{s}$-level sets. Symbols: Monte Carlo simulations results. The area below the thick blue line for $\frac{\gamma}{s}=0$ is not in the phase space of pumped branching processes.}
\label{fig_SpikeMomentsMapVerification}
\end{center}
\end{figure}

In Fig.~\ref{fig_SpikeMomentsMapExpData}\textbf{A}, data points from experiments were plotted onto a map of spike moments, which was then transformed into a map in $\frac{r}{s}$-$\frac{\gamma}{s}$-space. In this section, the analytics in this figure are verified using simulations which are shown in Fig.~\ref{fig_SpikeMomentsMapVerification}. 

Although the formulas in Eqs.~\eqref{eq-spike-waiting-time-from-fixed-state},~\eqref{eq-spike-waiting-time} for the level sets are analytical, calculating specific values requires special functions and high precision numerical evaluation, which makes accessing this map difficult for areas in the top left of Fig.~\ref{fig_SpikeMomentsMapVerification}, i.e. for small $X$ and large $Y$. A program for calculating moments of inter-spike intervals induced by pumped branching processes is available at \cite{Pausch2020b}. The area below the thick blue line for $\frac{\gamma}{s}=0$ is not in the phase space of the spike time series induced by pumped branching processes.

\section{Approximation Errors}\label{sec-approx-error}
\begin{figure}
\begin{center}
\includegraphics[width=\columnwidth]{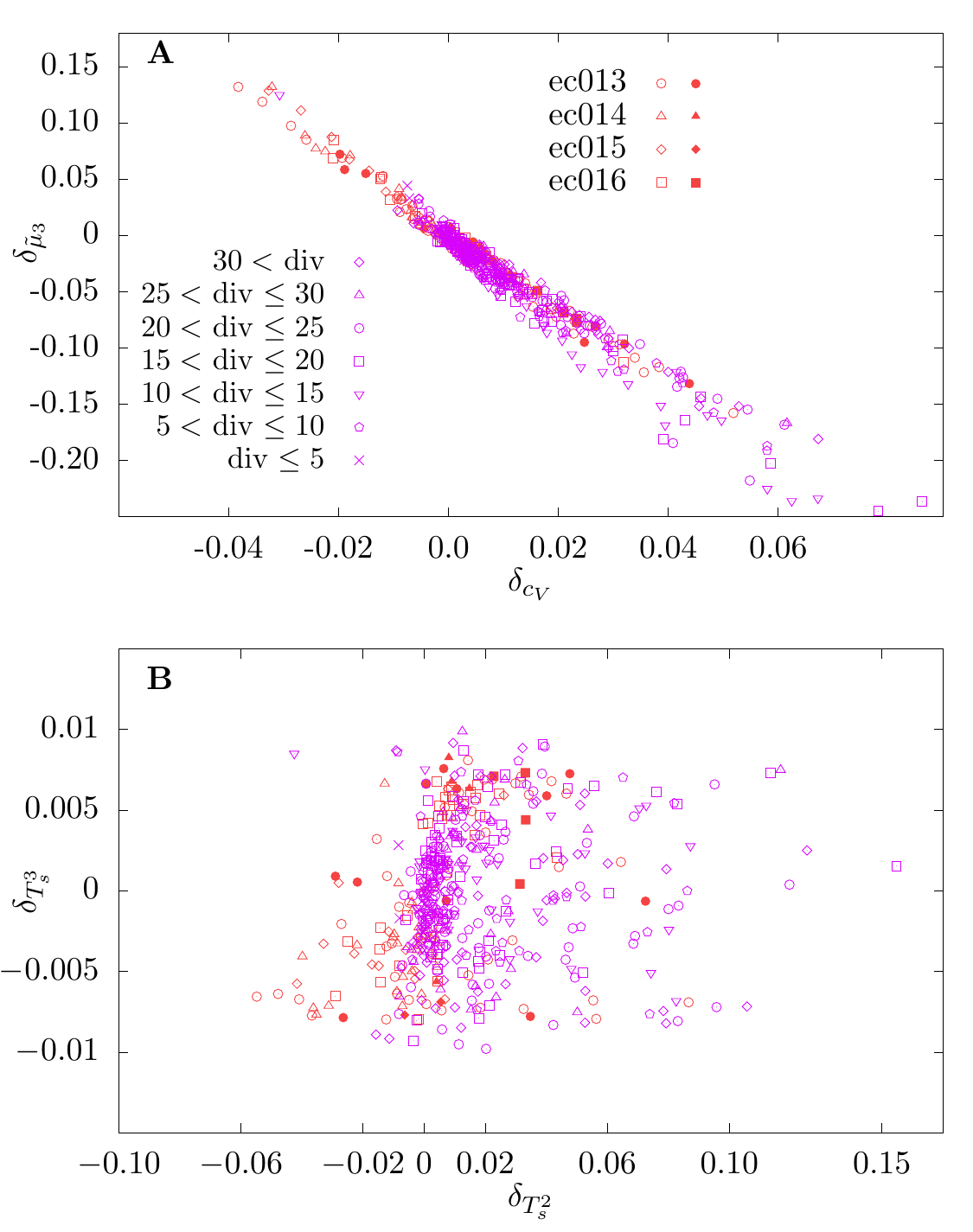}
\caption{Approximation errors for \textit{in vivo} data sets from \cite{Buzsaki2009} and \textit{in vitro} data sets from \cite{Wagenaar2006}. \textbf{A}: Signed approximation errors of the coefficient of variation $c_V$ and skewness $\tilde\mu_3$. \textbf{B}: Signed approximation error of the 2nd and 3rd moments of the inter-spike interval $T_s$. All approximation errors appear to indicate that the estimation is unbiased over all data sets.}
\label{fig_Approximation_Error}
\end{center}
\end{figure}

The moment-ratio map in Fig.~\ref{fig_SpikeMomentsMapExpData}\textbf{A} is used to identify values for the degree of criticality  $\frac{r}{s}$ and relative spontaneous creation $\frac{\gamma}{s}$ of experimental data, which then are used to identify the time scale $s$ using the first moment of inter-spike interval $T_s$ alone. There are many reasons why these parameter values have errors. Firstly, the pumped branching model and its induced spike process is an idealization of the true process in the brain. Second, experimental measurements always contain errors due to used instruments and choice of data processing.  Furthermore, the identification of the parameters is done algorithmically using a gradient descent method and has inherent inaccuracies that are chosen to limit run time. In addition, it is based on calculation using 128bit double precision. Although 128bit double precision seems to be an unusual restriction, it is a serious one in this case because close to criticality, the probabilities for the system to be in a specific state $N=n$ tend to zero. This implies that the calculation of the moments of $T_s$ involves summing over many near-zero probabilities and thus the 128bit double precision becomes a limiting factor and a source of errors. 

The effect of all of these sources of errors result in approximation errors of the true inter-spike interval distribution. The magnitude of these errors determines the accuracy and predictive power of the model. For an observable $\mathcal{O}$, the signed approximation error is defined as \begin{align}\delta_\mathcal{O}=\frac{\mathcal{O}_\text{data}-\mathcal{O}_\text{predicted}}{\mathcal{O}_\text{predicted}}.
\end{align} 
Fig.~\ref{fig_Approximation_Error} shows the signed approximation errors of the coefficient of variation $c_V$ and the skewness $\tilde\mu_3$ (Panel \textbf{A}), as well as the 2nd and 3rd moments of the inter-spike interval (Panel \textbf{B}). Except for the skewness, the approximation errors lie in a range of a few percent and seem to be unbiased over all data sets. As the skewness and coefficient of variation are both depending on the mean and standard deviation, a systematic relationship between their errors is not surprising.  

\section{Description of data collection procedure}\label{app-data-description}
The first batch of data sets \cite{Buzsaki2009} was downloaded from crcns.org, where it is called data set hc-2. The following is a quote from the description document of the data set: 

'Three male Long-Evans rats (rat ID; ec13, ec14, ec16, 250-400 g) were implanted with a 4-shank or 8-shank silicon probe in layer CA1 of the right dorsal hippocampus. The individual silicon probes were attached to respective micromanipulators and moved independently. Each shank had 8 recording sites (160 $\mu$m2 each site; 1-3 M$\Omega$ impedance). These recordings sites were staggered to provide a two-dimensional arrangement (20 $\mu$m vertical separation; Fujisawa et al., 2008). The shanks were aligned parallel to the septo-temporal axis of the hippocampus (45 degrees parasagittal), positioned centrally at anteroposterior=3.5 mm from bregma and mediolateral=2.5 mm from midline. Two stainless steel screws inserted above the cerebellum were used as indifferent and ground electrodes during recordings. All protocols were approved by the Institutional Animal Care and Use Committee of Rutgers University.'

The second batch of data sets \cite{Wagenaar2006} was downloaded from \url{neurodatasharing.bme.gatech.edu/development-data/}. All data sets labelled as 'dense' are used in this article. The following is paraphrasing and a short summary of the full description in \cite{Wagenaar2006}:
For the data collection, multi-electrode arrays were used with 59 electrodes of 30$\mu$m diameter. The electrodes were arranged in a square grid with 200$\mu$m spacing between them, measured center-to-center. Neurons and glia were obtained from rat embryos. Quote: "[...] timed-pregnant Wistar rats were sacrificed using CO$_{-32}$ inhalation, according to NIH approved protocols, at day 18 of gestation. Embryos were removed and decapitated, and the anterior part of the cortices (including somatosensory, motor, and association areas) were dissected out. At the rostral edge, the boundary with the olfactory bulb was used as a landmark; at the caudal edge, the third ventricle and the boundary with the lateral horn of the hippocampus were used as landmarks. Striatum and hippocampus were not included. Cortices from several embryos from the same litter were combined, [...]". For the maintenance of the cultures, quote: "Half of the medium was replaced approximately every five days in most experiments [...]. To test whether feeding schedule affected activity, all medium was replaced every seven days in some experiments (N = 3). This did not result in significantly different activity patterns compared with sister cultures. Feeding always took place after the day's recording session, to allow at least 12 hours for transient effects to disappear before the next recording."

\bibliography{references}
\end{document}